\documentclass[a4paper,twoside]{article}
\newcommand{\affil}[1]{$^{\rm #1}$}
\usepackage[authoryear]{natbib}
\bibpunct{(}{)}{;}{a}{}{,}
\usepackage{graphicx}
\date{} 
\def\kms{km${\rm s}^{-1}$}

\def\arcmin{\hbox{$^\prime$}}
\def\arcsec{\hbox{$^{\prime\prime}$}}

\def\ha{H$\alpha$}
\def\hb{H$\beta$}
\def\hg{H$\gamma$}

\def\NII{[N\,\textsc{ii}]}
\def\NI{[N\,\textsc{i}]}
\def\OI{[O\,\textsc{i}]}
\def\OII{[O\,\textsc{ii}]}
\def\OIII{[O\,\textsc{iii}]}

\def\OVI{O\,\textsc{vi}}
\def\SII{[S\,\textsc{ii}]}
\def\HII{H\,\textsc{ii}}
\def\HeII{He\,\textsc{ii}}
\def\HeIII{He\,\textsc{iii}}
\def\FeII{[Fe\,\textsc{ii}]}

\def\p0{\phantom{0}}
\def\lessim{\raise-.5ex\hbox{$\buildrel<\over\sim$}}
\def\grtsim{\raise-.5ex\hbox{$\buildrel>\over\sim$}}

\title{\large\bf\flushleft Planetary Nebulae: Observational Properties, Mimics, and Diagnostics}
\author{\parbox{\textwidth}{\flushleft
\vspace{-0.5cm}
{\it David J. Frew\affil{A,C,D}, and Quentin A. Parker\affil{A,B}}\\
\vspace{0.4cm}
{\small \affil{A}\,Department of Physics, Macquarie University, North Ryde, NSW 2109, Australia}\\
{\small \affil{B}\,Anglo-Australian Observatory, Epping, NSW 1710, Australia}\\
{\small \affil{C}\,Perth Observatory, Bickley, WA 6076, Australia}\\
{\small \affil{D}\,Email: dfrew@science.mq.edu.au}}}
\begin{document}
\twocolumn[
\begin{changemargin}{.8cm}{.5cm}
\begin{minipage}{.9\textwidth}
\vspace{-1cm}
\maketitle
%
%
\small{\bf Abstract:}
The total number of true, likely and possible planetary nebulae (PN) now known in the Milky Way is nearly 3000, double the number known a decade ago.  The new discoveries are a legacy of the recent availability of wide field, narrowband imaging surveys, primarily in the light of H-alpha.  In this paper, we summarise the various PN discovery techniques, and give an overview of the many types of objects which mimic PN and which appear as contaminants in both Galactic and extragalactic samples.   Much improved discrimination of classical PN from their mimics is now possible based on the wide variety of high-quality multiwavelength data sets that are now available.  We offer improved taxonomic and observational definitions for the PN phenomenon based on evaluation of these better diagnostic capabilities.  However, we note that evidence is increasing that the PN phenomenon is heterogeneous, and PN are likely to be formed from multiple evolutionary scenarios.  In particular, the relationships between some collimated symbiotic outflows and bipolar PN remain uncertain.



\medskip
\medskip
\end{minipage}
\end{changemargin}
]
\small

\section{Introduction}\label{sec:PN_intro}

A classical planetary nebula (PN) is a shell of ionized gas, ejected from a low- to intermediate-mass star ($\sim$1 to 8 M$_{\odot}$) towards the end of its life.  A PN is produced at the end of the asymptotic giant branch (AGB) phase, as the red giant ejects its outer envelope in a final stage of copious mass loss termed the `superwind'.  After the envelope ejection, the remnant core of the star increases in temperature before the nuclear burning ceases and the star quickly fades, becoming a white dwarf (WD).  The high temperature of the central star (CS) causes the previously ejected material to be ionized, which becomes visible as a PN, its shape sculpted by the interaction between the old red giant envelope and a tenous, fast wind from the hot CS (Kwok, Purton \& Fitzgerald 1978). 

PN\footnote{For simplicity, we use PN as an abbreviation for both singular and plural forms.} are an important, albeit brief ($\leq$10$^5$ yr), evolutionary phase in the lifetimes of a significant fraction of Milky Way stars.  They are an important tool in our understanding of the physics of mass loss for intermediate-mass stars, the chemical enrichment of our Galaxy, and in turn, its star formation history.  PN are ideal test particles to probe the dynamics of the Milky Way and are amongst the best kinematic tracers in external galaxies.  The PN luminosity function is also a powerful extragalactic distance indicator (e.g. Ciardullo 2010, this issue).

The total number of currently known Galactic PN is nearly 3000, which includes all known {\sl confirmed} PN from the catalogues of Acker et al. (1992, 1996) and Kohoutek (2001), published PN found since 2000, including $\sim$1250 from the recent MASH catalogues (Parker et al. 2006; Miszalski et al. 2008), additional discoveries from the IPHAS survey (Viironen et al. 2009a, 2009b), the new faint PN found as part of the ongoing Deep Sky Hunters project (Kronberger et al. 2006; Jacoby et al. 2010, this issue), plus a modest number of additional PN gleaned from the literature.  

It is beyond the scope of this paper to give an exhaustive summary of our current state of knowledge concerning PN.  For reviews on different aspects of the PN phenomenon, the reader is referred to the works of Pottasch (1992), Balick \& Frank (2002), and De Marco (2009), and the monograph by Kwok (2000).

\section{Planetary Nebulae: a working definition}\label{sec:PN_definition}

There has been considerable controversy in the past over the working definition of a PN, based in part on the differing morphological, spectroscopic, and physical or evolutionary criteria used for classification, which have changed over the two centuries since their discovery.  

It is commonly argued that the PN moniker has a distinct physical or evolutionary meaning, and should be restricted to the ionized gaseous shell ejected at the end of the AGB phase, either by a single star, or as part of a common-envelope ejection (De Marco 2009, and references therein).   This is an important point, as in many symbiotic systems, which are often confused with PN, the gas is thought to be donated by a {\em companion} giant, and not the precursor of the hot WD.  The WD  provides the ionizing photons in these systems (Corradi 2003), and may have passed through a PN phase long before.

Surprisingly, a consensus taxonomic definition for PN remains elusive, even at present.  Indeed, there may be enough diversity in the PN phenomenon that a pure definition may be unattainable (see \S\ref{sec:hetero}, below).  Nevertheless, based on an overview of the literature, extensive experience in the compilation of the MASH catalogues, as well as an analysis of a large volume-limited PN census now containing $\sim$400 PN centred on the Sun (e.g. Frew \& Parker 2006; Frew 2008; Frew et al. 2010, in preparation), we offer the following phenomenological definition (cf. Kohoutek 2001).  

We define a PN as an ionized emission nebula ejected from what is now the hot central star (CS), which exhibits the following observational and physical characteristics:

\begin{itemize}
\item A round or axisymmetric (elliptical or bipolar) shape, sometimes with multiple shells or outer haloes. Point-symmetric microstructures (Gon\c{c}alves, Corradi \& Mampaso 2001) are more common in higher-surface brightness PN, while many evolved PN are perturbed through interaction with the interstellar medium (ISM).

\item A primarily photoionized emission-line spectrum characterised by recombination lines of hydrogen and helium as well as various collisionally-excited forbidden lines of heavier elements such as O, N, C, Ne, Ar and S, which are prominent in the UV, optical and IR regions.  Signatures of shock excitation are seen in many objects, especially some bipolar PN with fast outflows, and PN strongly interacting with the ISM.  The [O\,{\sc iii}] lines at $\lambda\lambda$4959, 5007\AA\ are usually, but not always, the strongest emission lines in the optical, the exceptions being very low-excitation (VLE) PN with cool central stars (where the \ha\ line may dominate), some Type~I objects with very strong $\lambda\lambda$6548,6584\AA\ [N\,{\sc ii}] lines, and/or PN suffering moderate to heavy extinction.

\item A dereddened \ha\ surface brightness between the approximate empirical limits of log\,$S$(H$\alpha$) = +0.5 and $-$6.5 erg\,cm$^{-2}$s$^{-1}$sr$^{-1}$ (or $-$10.1 to $-$17.1 erg\,cm$^{-2}$s$^{-1}$arcsec$^{-2}$).

\item A reddening-corrected \OIII\ absolute magnitude, $M_{\rm 5007}$ between the empirical limits of $-$4.5 and +6.0 mag.  Note that some very young PN with cool CS may have negligible \OIII\ emission.

\item Thermal free-free (bremsstrahlung) emission in the radio spectrum.

\item Signature of cool dust ($T_{\rm dust}$ $\simeq$ 100--200\,K) in the infrared, in young to mid-aged PN.  Dust emission peaks at around 20\,$\mu$m for most PN.

\item Sometimes PAH emission at 8$\mu$m from the enveloping photo-dissociation region, but mostly in more compact objects.

\item A nebular radius, $r$ $\leq$ 2.5 pc, though the vast majority have $r$ $<$ 1.5 pc (though see Section~7, below). Note that the median radius of a volume-limited sample within 1\,kpc (dominated by old PN) is $\sim$0.6 pc.   Young PN (and pre-PN) typically have radii $<$ 0.05\,pc.  Many young and mid-aged PN are surrounded by extended AGB haloes, which typically are a factor of $\sim$10$^3$ fainter than the main PN shell (Corradi et al. 2003).

\item An ionized mass roughly between the empirical limits of 0.005 and 3\,$M_{\odot}$, with a median value of 0.5~$\sqrt{\epsilon}$\,$M_{\odot}$, where $\epsilon$ is the volume-filling factor.  These limits differentiate PN from both low-mass nova shells and high-mass circumstellar shells around massive Population~I stars.   Recent observations have shown that the ionized mass of some symbiotic outflows approaches 0.1\,$M_{\odot}$, so there is substantial overlap between the classes (see Santander-Garcia et al. 2008).

\item An electron density between 10$^0$ and 10$^5$ cm$^{-3}$, which is generally less than the densities seen in symbiotic stars (10$^6$ to $>$10$^{10}$ cm$^{-3}$).

\item A shell expansion velocity typically between 10--60 \kms\ though some strongly bipolar PN can have considerably higher expansion velocities (up to 300 \kms) along the major axis.

\end{itemize}

By definition, a PN surrounds a hot, low-mass CS, which may in fact be off-centre in some old examples due to an asymmetric ISM interaction (see Wareing 2010, and Sabin et al. 2010, in this issue).  The CS has a temperature of at least 20,000\,K (up to $\sim$250,000\,K), and a mass between the empirical limits of $\sim$0.55 and 0.9\,$M_{\odot}$.  It is possible that PN surrounding lower mass CS are formed via a common-envelope process, while PN around higher mass CS (up to the Chandrasekhar limit at $\sim$1.35\,$M_{\odot}$) are not observed due to selection effects, i.e. very rapid evolution of the CS.  

The observed absolute visual magnitudes of CS are in the range, $-2.5$ $\geq$ $M_{V}$ $\leq$ +7.5.  Corresponding surface gravities are generally in the range, log\,$g$\,$\simeq$ 2.5--7.5 cms$^{-2}$.  Gravities are lower (0.5--2.5\,cms$^{-2}$) in the immediate progenitors of PN, the post-AGB stars (for a review, see Van Winckel 2003).

Spectral types for the majority of CS range from OB and sdO through to DAO and DA.  Up to 20\% of CS are H-deficient, including those where Wolf-Rayet (WR) features are present --- these are denoted [WR] to differentiate them from their Population I cousins.  Almost all belong to the [WC]/[WO] sequence except for the rare [WN] objects N\,66 in the LMC (Hamann et al. 2003) and PM~5 in the Galaxy (Morgan, Parker \& Cohen 2003).  However, further work is needed to completely rule out the possibility that this latter object is a Population~I WR ring nebula.  New multiwavelength observations of N\,66 are needed regardless of other arguments, considering that over the past twenty years it has exhibited rapid changes in its spectrum. N66 seems to have developed its WR features over the course of only three years (Pe\~na et al. 1995), suggesting that the [WN] phase might be a temporary phenomenon in intermediate-mass H-deficient stars. 

Other H-deficient stars include the weak emission-line stars (\emph{wels}), and the PG~1159 types, which are thought to be the progeny of the [WC] and [WO] stars (for reviews, see Marcolino \& de Ara\'ujo 2003; Werner \& Herwig 2004; and De Marco 2008).

\section{PN Detection Techniques}\label{sec:detection}

As befits the large range in apparent size, morphology, integrated flux, surface brightness, excitation class,  central star magnitude, and evolutionary state shown by individual PN, a large and diverse range of detection methods has been used in their discovery.  Some of the brightest and best-known PNe were found over two centuries ago at the telescope eyepiece by pioneering observers like Charles Messier and William Herschel.  Herschel in particular assigned nebulous objects to the PN class purely on morphological grounds.   The various methods used to discover PN over the last hundred years or so are briefly summarised below, presented as a guide to the reader.\\


\subsection{Spectroscopic discoveries}

Many early PN discoveries were made spectroscopically, initially by visual methods, but primarily from objective-prism photographic plates.  The middle decades of the twentieth century provided a vast increase in the number of new PN as photographic surveys reached greater depth, typified, for example, by the work of Minkowski (1946), Wray (1966), and Henize (1967).  This technique has been particularly applicable to the Galactic bulge, which contains large numbers of compact, relatively high-surface brightness PN.\\

\subsection{Visual Inspection of Optical Imagery}

Numerous evolved PN were discovered from painstaking visual scrutiny of large numbers of broadband POSS~I and ESO/SERC Schmidt plates and films (e.g. Kohoutek 1963; Abell 1966; Longmore 1977; Dengel, Hartl \& Weinberger 1980; Lauberts 1982; Kerber et al. 1998, and references therein).  More recently, Whiting, Hau \& Irwin (2002) and Whiting et al. (2007) found several faint PN candidates from the POSS-II and ESO/SERC surveys while searching for nearby dwarf galaxies.  It should be emphasised that spectroscopy is required to confirm the PN candidates found in this way.

A vast increase in known PN numbers has occurred in the last decade due to the advent of deep, narrowband, high-resolution surveys of the Galactic plane in the light of \ha.  The first of these was the photographic Anglo-Australian Observatory / UK Schmidt Telescope (AAO/ UKST) \ha\ survey of the Southern Galactic plane.  The photographic films have been completely digitized and are now available online as the SuperCOSMOS \ha\ Survey (SHS; Parker et al. 2005).  This survey has provided the source material for the Macquarie/ AAO/ Strasbourg \ha\ (MASH) catalogues of Galactic PN of Parker et al. (2006; MASH-I) and Miszalski et al. (2008; MASH-II).  These contain a total of $\sim$1250 true, likely and possible PN, which represented a near doubling of the PN population known at that time.

More recently, discoveries have come from direct inspection of narrow-band CCD survey imagery.  For example, many highly-evolved PN candidates (e.g. Sabin et al. 2010) are being found on \ha\ mosaic images from the Isaac Newton Telescope Photometric H$\alpha$ Survey (IPHAS, Drew et al. 2005), which covers the northern Galactic plane. 

Finally, we should mention the PN discovered directly from broadband Digitized Sky Survey (DSS) digital images, including $\sim$100 candidates found from systematic visual scans of  online images conducted by a team of amateur astronomers (see Kronberger et al. 2006; Jacoby et al. 2010, these proceedings).  These recent finds hint that there are still numerous faint PN remaining to be discovered on broadband surveys outside  the zones covered by the AAO/UKST and IPHAS surveys.  \\

\subsection{Image Comparison Techniques}

Subtraction or quotient imaging has been applied to \emph{digitized} photographic survey plates and films; for example, Cappellaro et al. (1994) found three new PN and a Herbig-Haro object by comparing digitized POSS red and infrared (IR) plates.  This procedure works because the $I$-band filter passes relatively weak emission lines compared to the $R$-band which includes the bright \ha, \NII\ and \SII\ lines.   

For the MASH-II survey, Miszalski et al. (2008) systematically searched the SHS using blocked-down (factor 16$\times$)  {\sl quotient} images, generated from scanned \ha\ and matching broadband-red photographic data.  Candidates were then examined using combined RGB images made from the \ha, $SR$ and $B$-band SuperCOSMOS data at full resolution to help ascertain their nature.  

Comparison of deep on-band and off-band CCD imaging (Beaulieu, Dopita \& Freeman 1999; Boumis et al. 2003, 2006) has found additional PN in the Galactic bulge.  This technique has been used extremely successfully by Jacoby and co-workers for the discovery of numerous extragalactic PN (e.g. Jacoby et al. 1990; Jacoby \& De Marco 2002).  On the other hand, continuum-subtracted \ha\ images from the Southern \ha\ Sky Survey Atlas (SHASSA; Gaustad et al. 2001) were used to search for new large Galactic PN, but with only minimal success (Frew, Madsen \& Parker 2006).

Additionally for the MASH-II survey, the SHS parameterised object (IAM) data was mined for PN by investigating colour-magnitude plots and looking for outliers (see Miszalski et al. 2008, for further details).  More recently, the IPHAS consortium has discovered numerous compact PN candidates (Viironen et al. 2009a, 2009b) using a range of colour-colour plots. 

Finally we note the PN discoveries in the Large Magellanic Cloud (LMC) by Reid \& Parker (2006 a,b).  These authors used a novel \ha/$SR$ colour merging technique to find 460 new PN from a deep, stacked \ha\ + red continuum map of the central 25 deg$^{2}$ region of the LMC.\\

\subsection{Discoveries at non-optical wavelengths}

Search techniques at infrared and longer wavelengths have the advantage of avoiding the worst effects of interstellar extinction, especially close to the Galactic bulge and plane.  A large number of PN candidates have been selected from their IRAS colours (e.g. Preite-Martinez 1988; Pottasch et al. 1988; van der Veen \& Habing 1988; Ratag et al. 1990; Van de Steene \& Pottasch 1993; Kistiakowsky and Helfand 1995; Garc\'ia-Lario et al. 1997).  However, there is always a question mark over these PN candidates until confirmatory spectra (e.g. Su\'arez et al. 2006) and high-resolution optical/IR or radio images are obtained, and the success rates have so far been modest. 

Image comparison techniques have also been undertaken in the near-IR.  Jacoby \& Van de Steene (2004) have conducted an on/off-band CCD survey of a region of the Galactic bulge in the light of [S\,{\sc iii}] $\lambda$9532, discovering 94 candidate PN.  This technique has the benefit of detecting very reddened PN. 

Mid-IR imagery will allow the detection of extremely reddened PN that are invisible at optical wavelengths.  Already, finds have been made from {\sl Spitzer} GLIMPSE data (e.g. Cohen et al. 2005;  Kwok et al. 2008;  Phillips \& Ramos-Larios 2008), while Ramos-Larios et al. (2009) are undertaking a search using a multiwavelength approach.  Cohen et al. (2010) are investigating a range of diagnostic tools to find more PN in the GLIMPSE dataset.  Lastly, Carey et al. (2009) and Flagey et al. (2009) have noted over 400 compact ($<$1\arcmin) ring, shell and disk-shaped sources in the Galactic plane at 24 $\mu$m in \emph{Spitzer} MIPSGAL images; many of these objects may turn out to be strongly reddened, high-excitation PN.

\section{PN Mimics}\label{sec:mimics}

A heterogeneous variety of contaminating emission-line objects have been a big problem in PN catalogues over the years, primarily because many new PN candidates were classified using only one or two criteria.  Either the emphasis was placed on morphology (e.g. Abell 1966), or on the object having a PN-like spectrum (e.g. Wray 1966).  With limited data sets available, it is understandable why a range of different astrophysical objects have turned up in published PN catalogues.  Similarly, PN have been found lurking in lists of \HII\ regions, reflection nebulae, and galaxies, and even three candidate globular clusters have turned out to be PN after spectroscopic analysis (Bica et al. 1995; Parker et al. 2006).

\begin{figure*}
\begin{center}
\includegraphics[scale=0.21,angle=0]{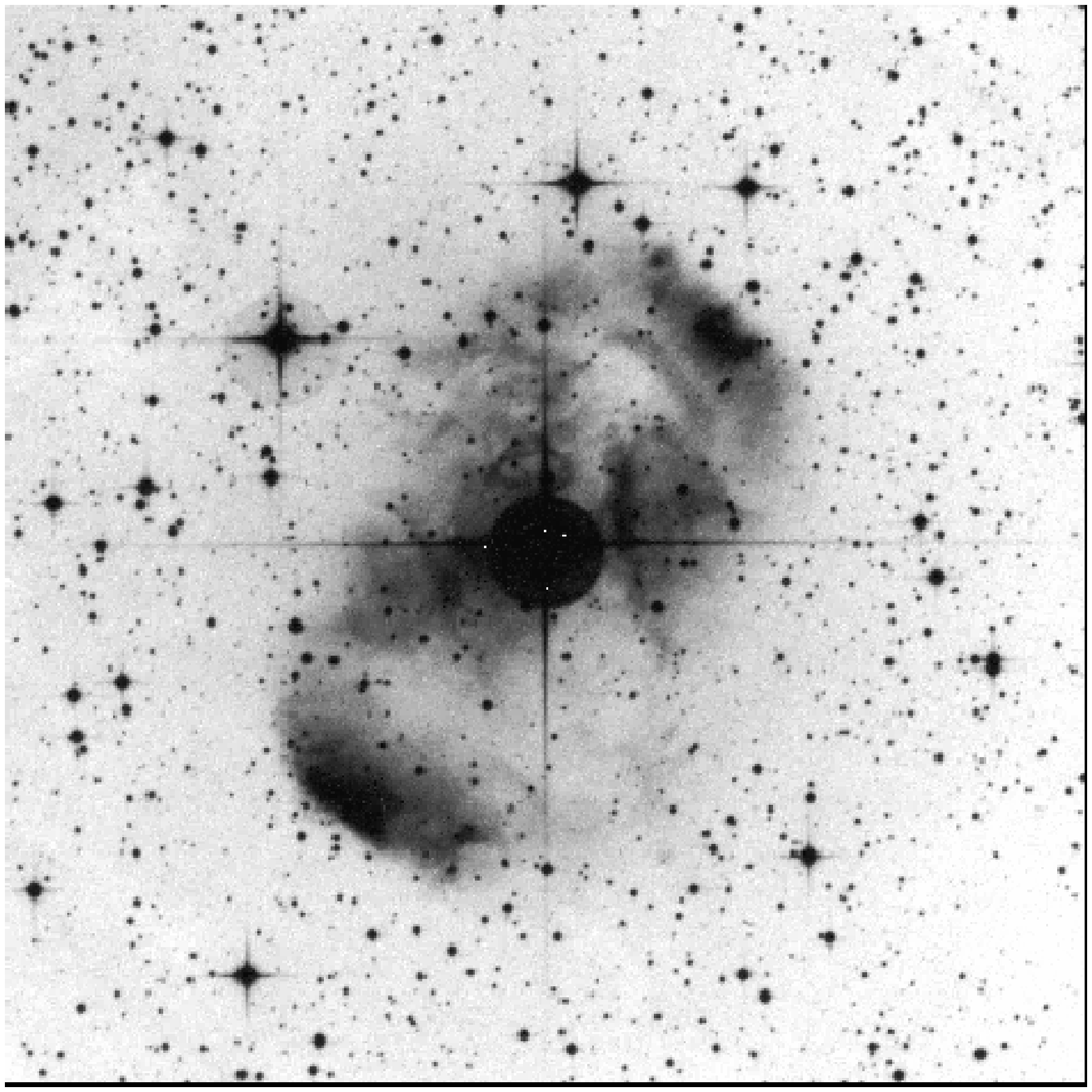}
\includegraphics[scale=0.21,angle=0]{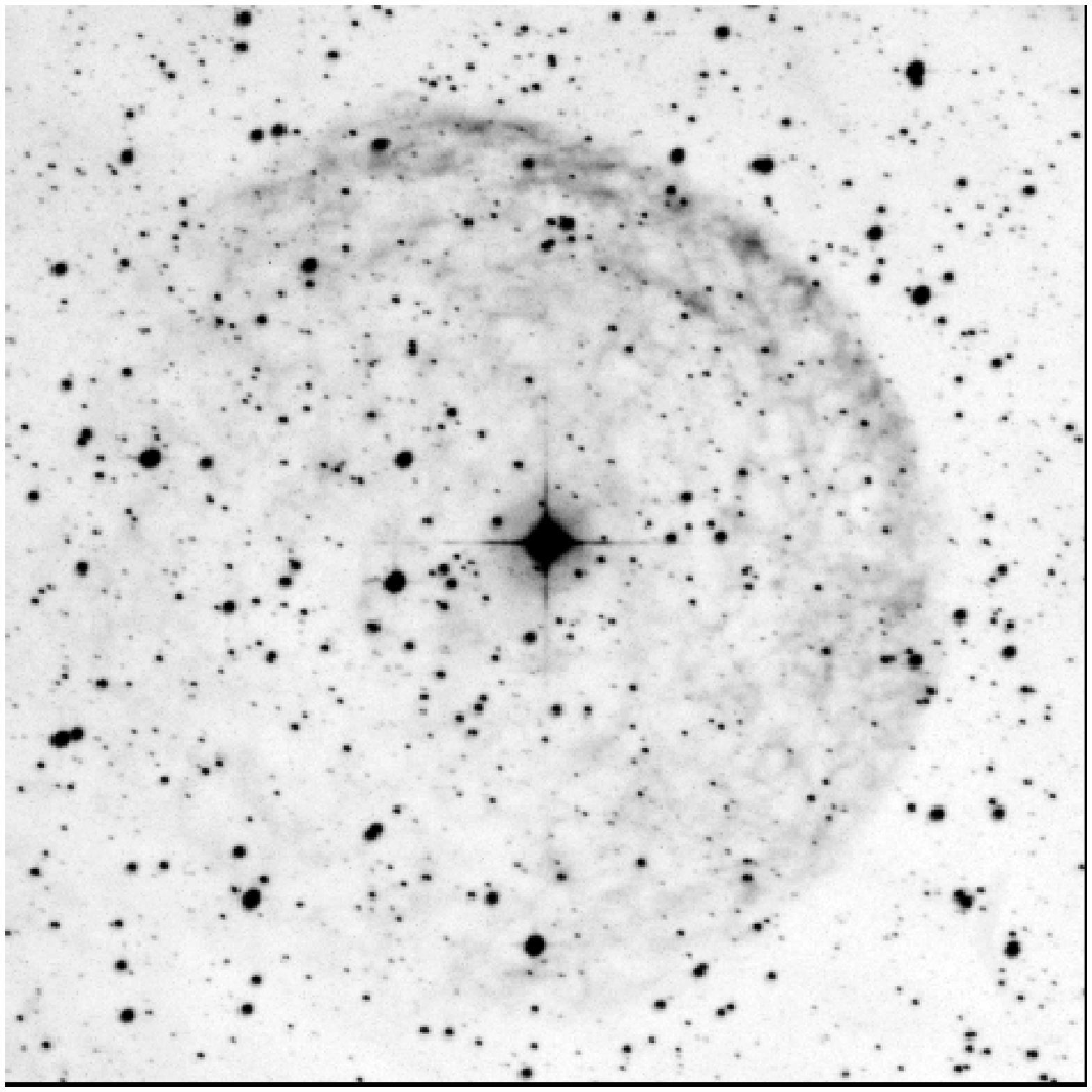}
\includegraphics[scale=0.21,angle=0]{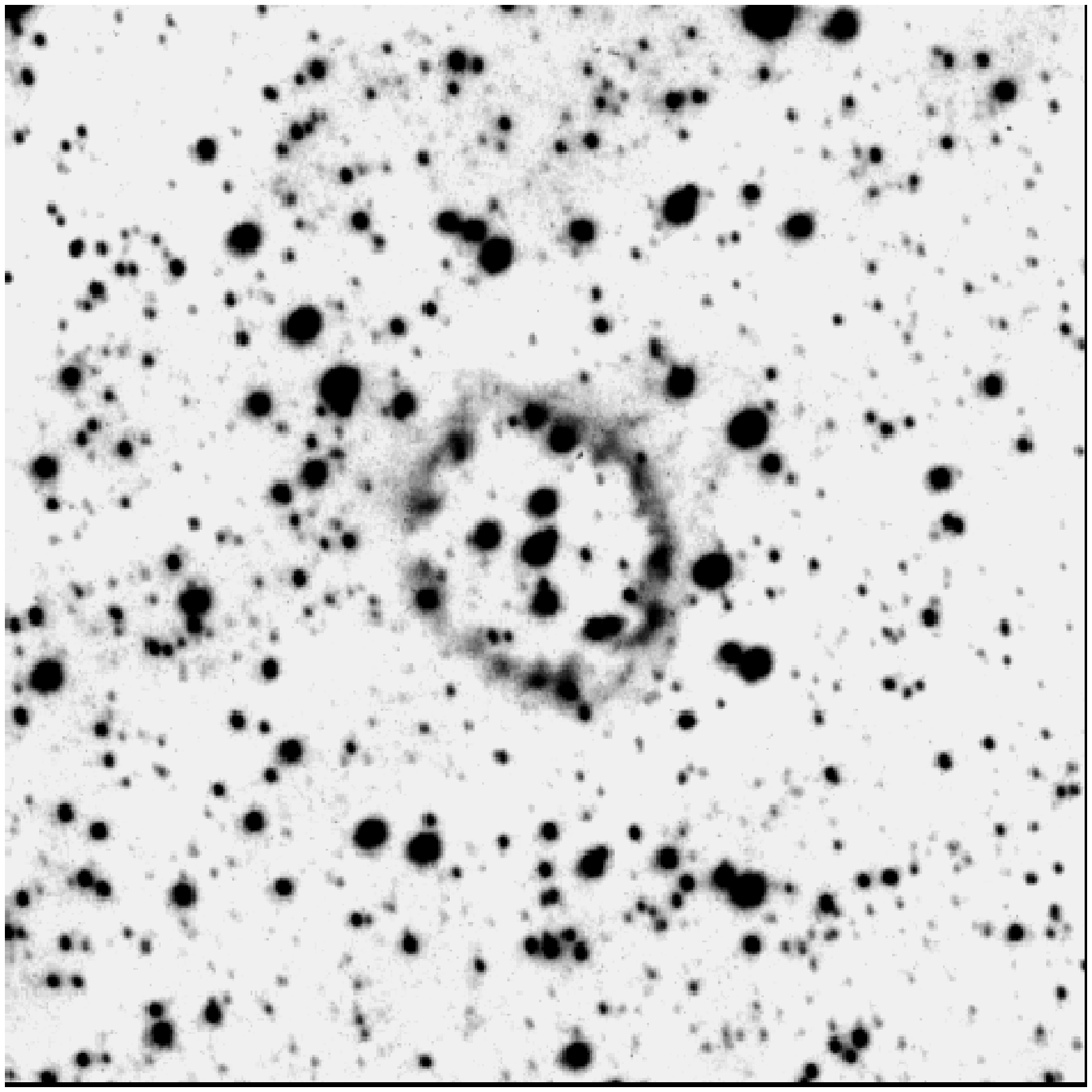}
\includegraphics[scale=0.21,angle=0]{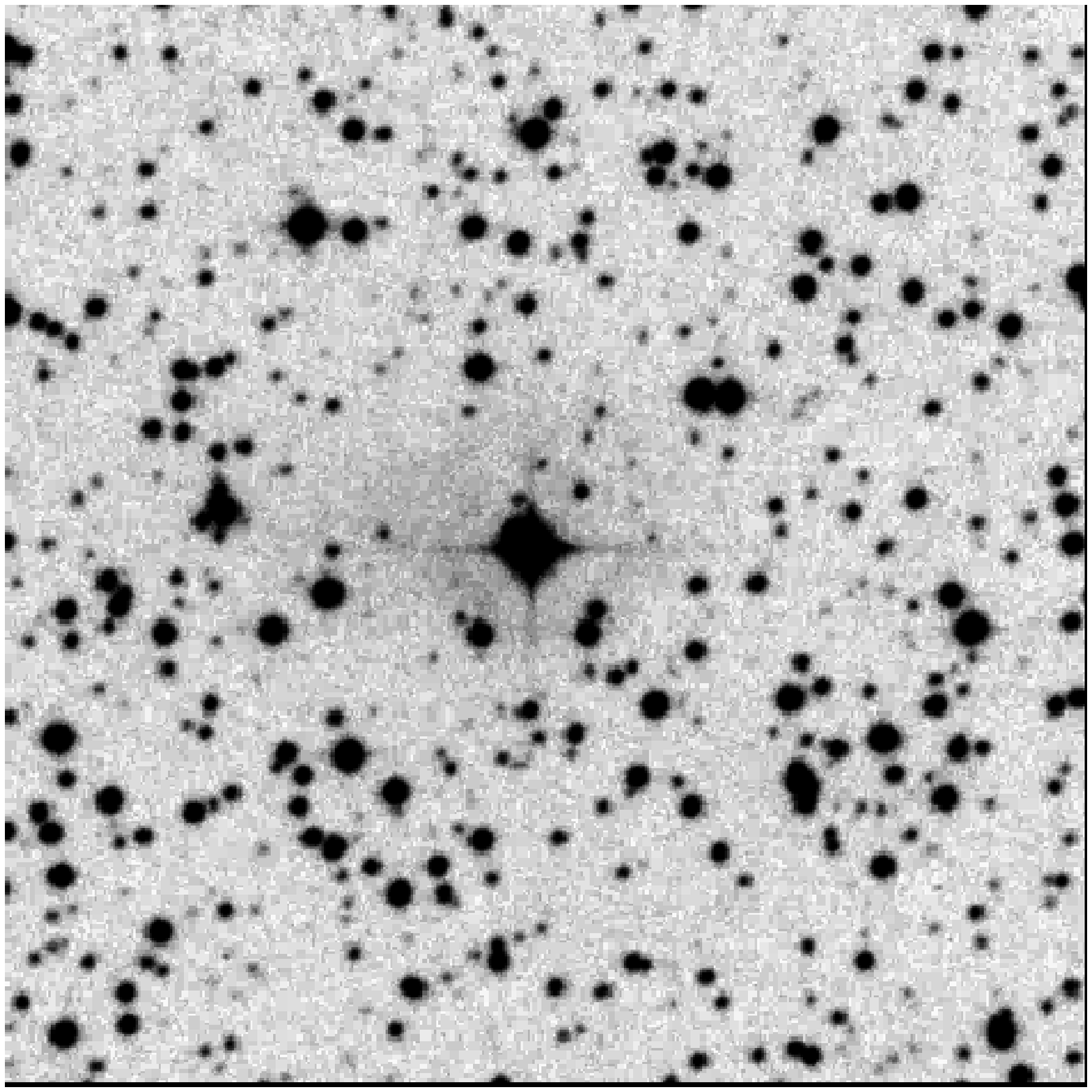}
\includegraphics[scale=0.21,angle=0]{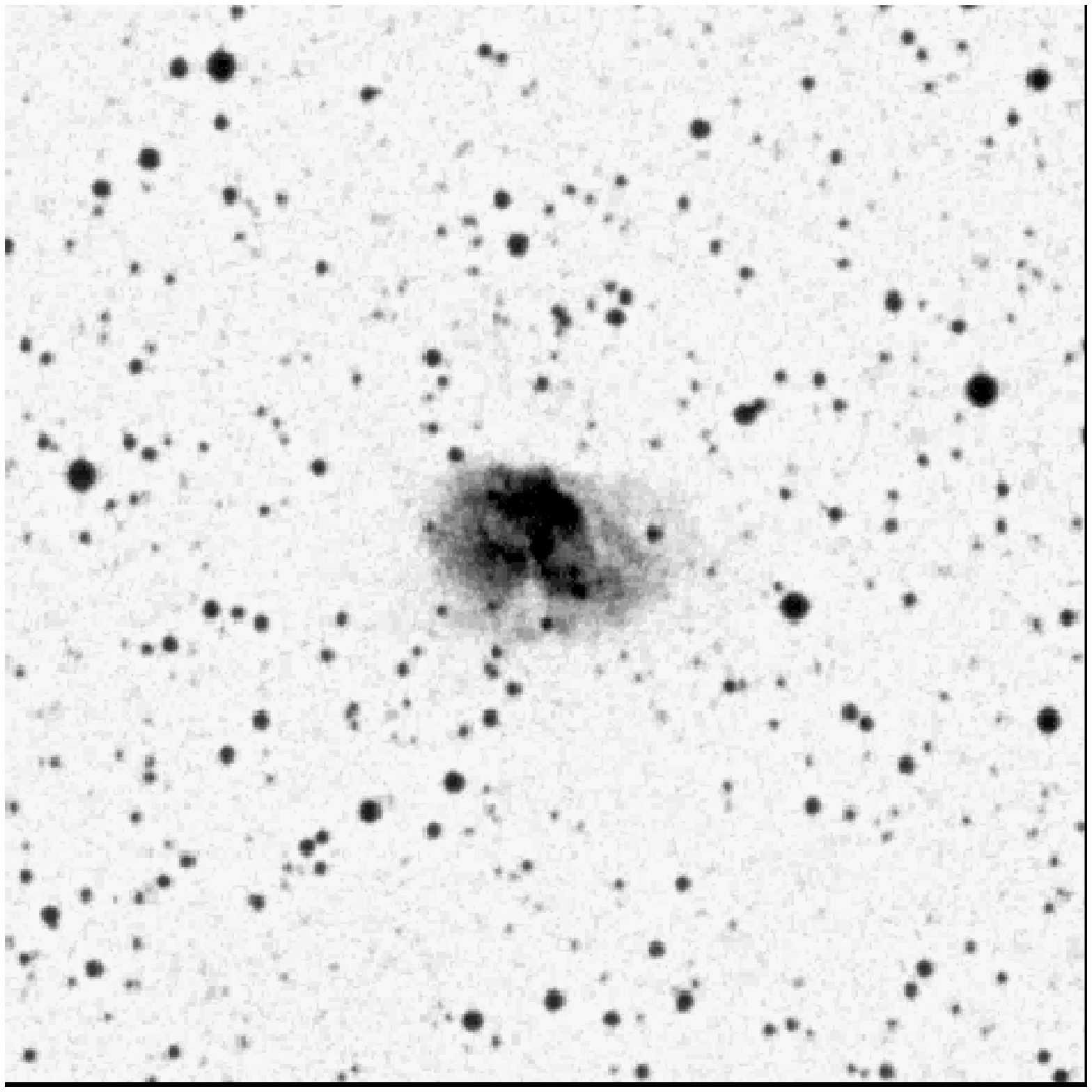}
\includegraphics[scale=0.21,angle=0]{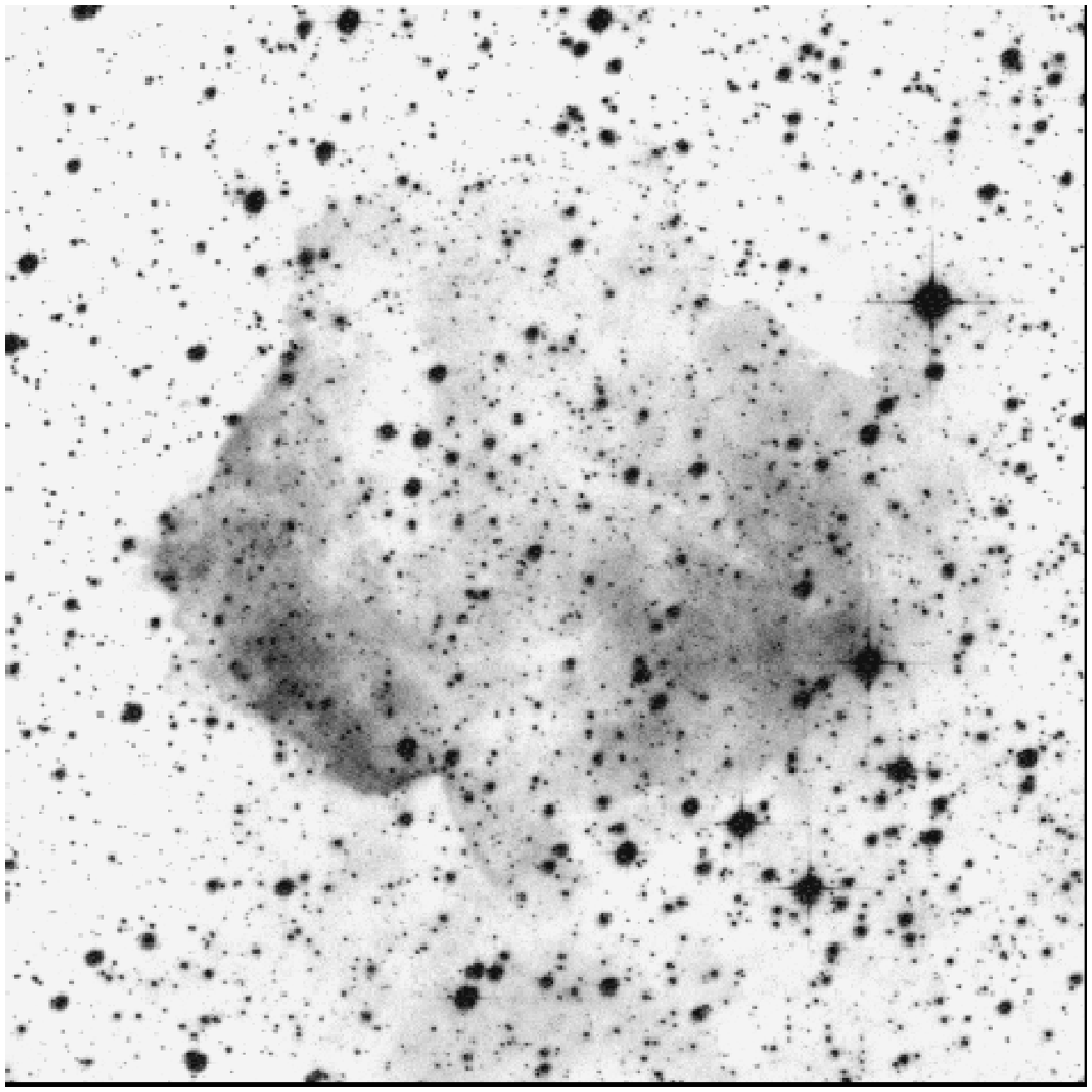}
\includegraphics[scale=0.06,angle=0]{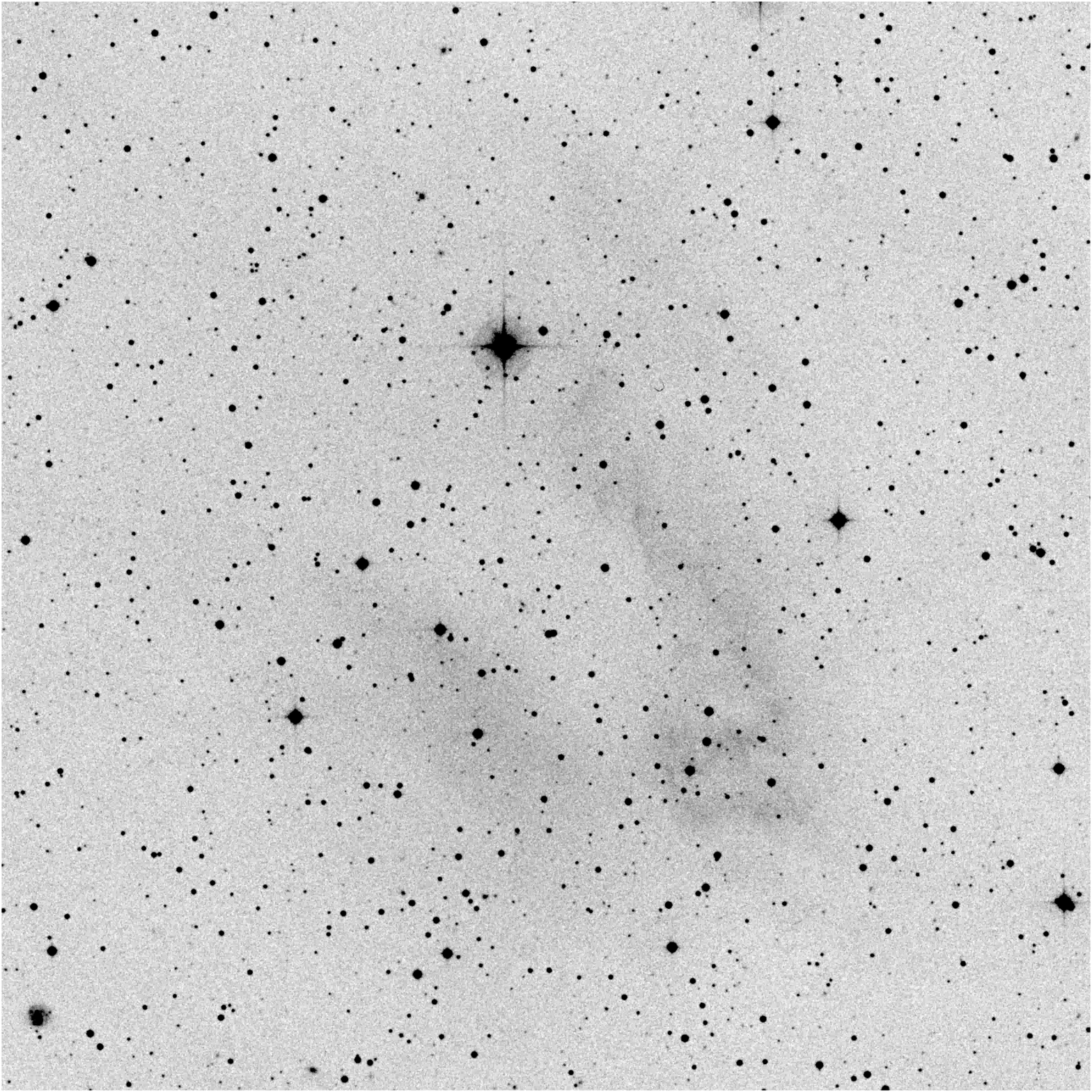}
\includegraphics[scale=0.395,angle=0]{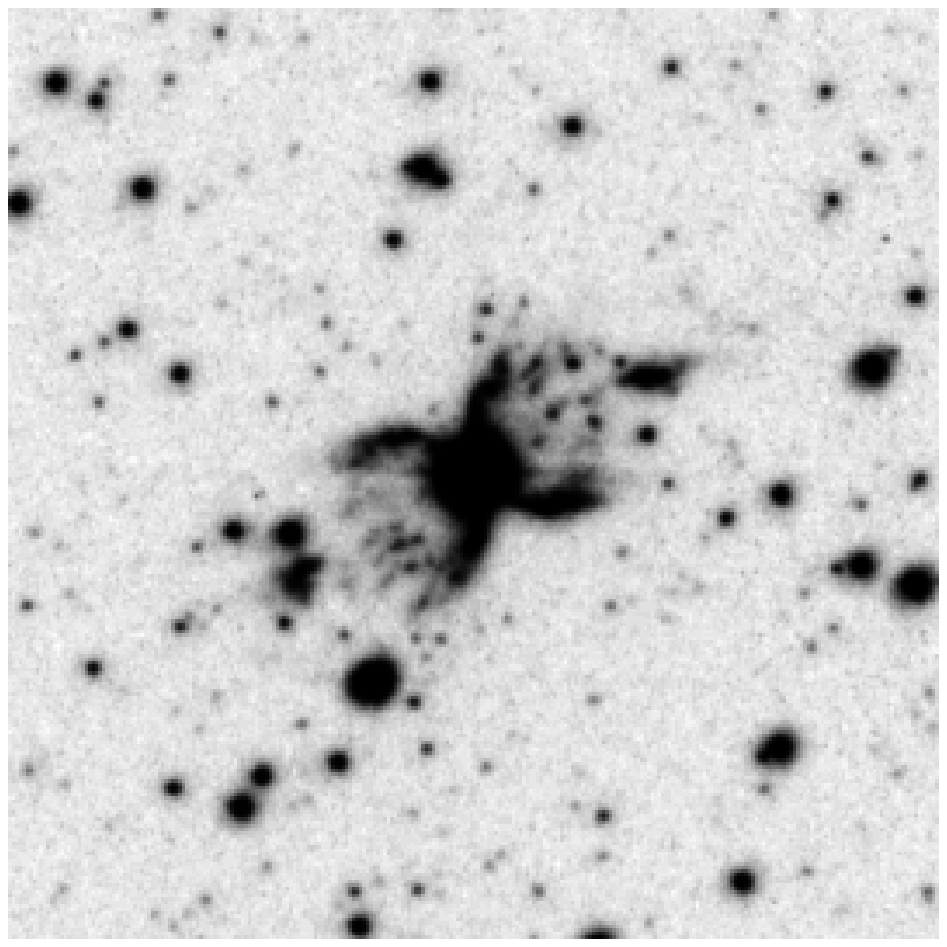}
\includegraphics[scale=0.30,angle=0]{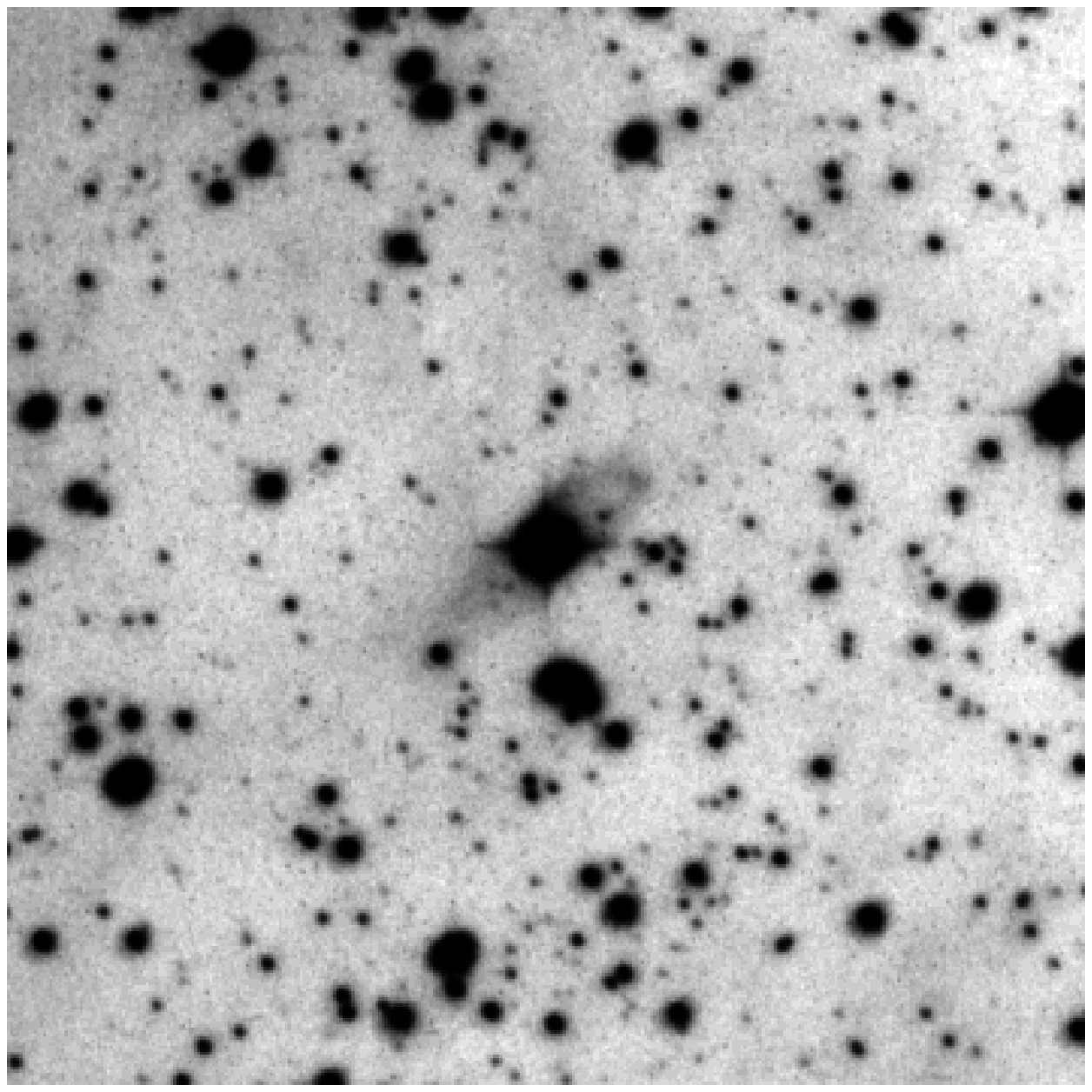}
\includegraphics[scale=0.21,angle=0]{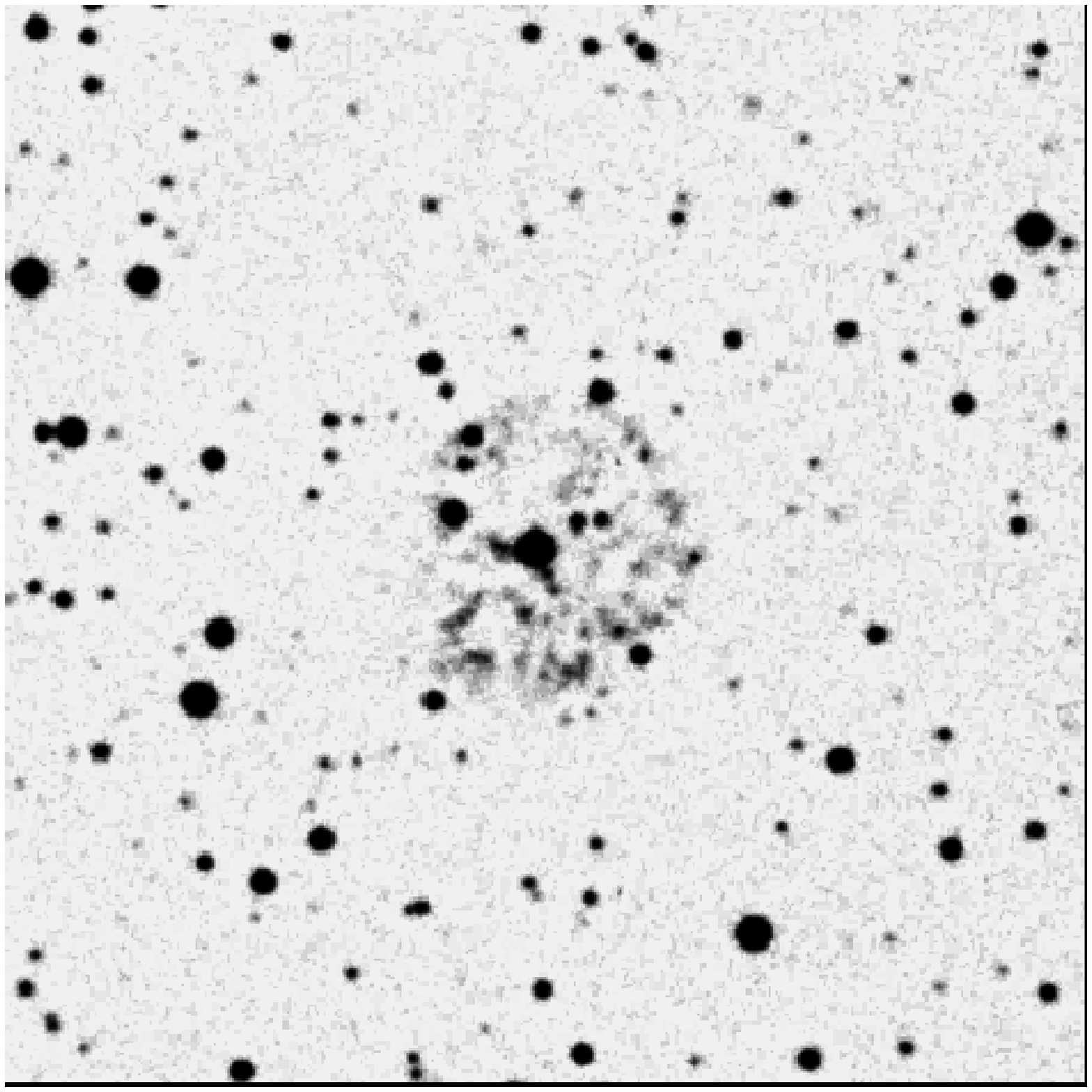}
\includegraphics[scale=0.21,angle=0]{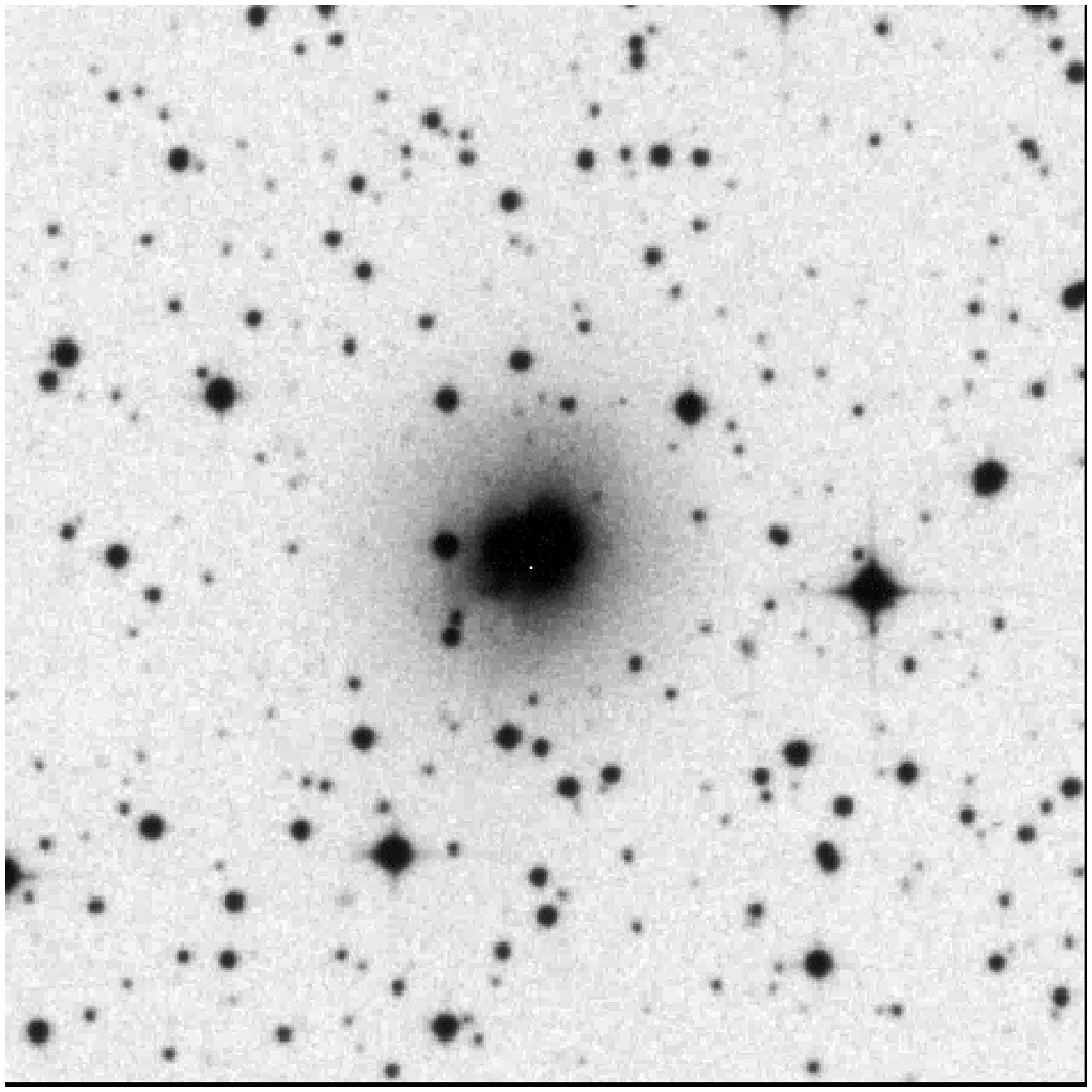}
\includegraphics[scale=0.21,angle=0]{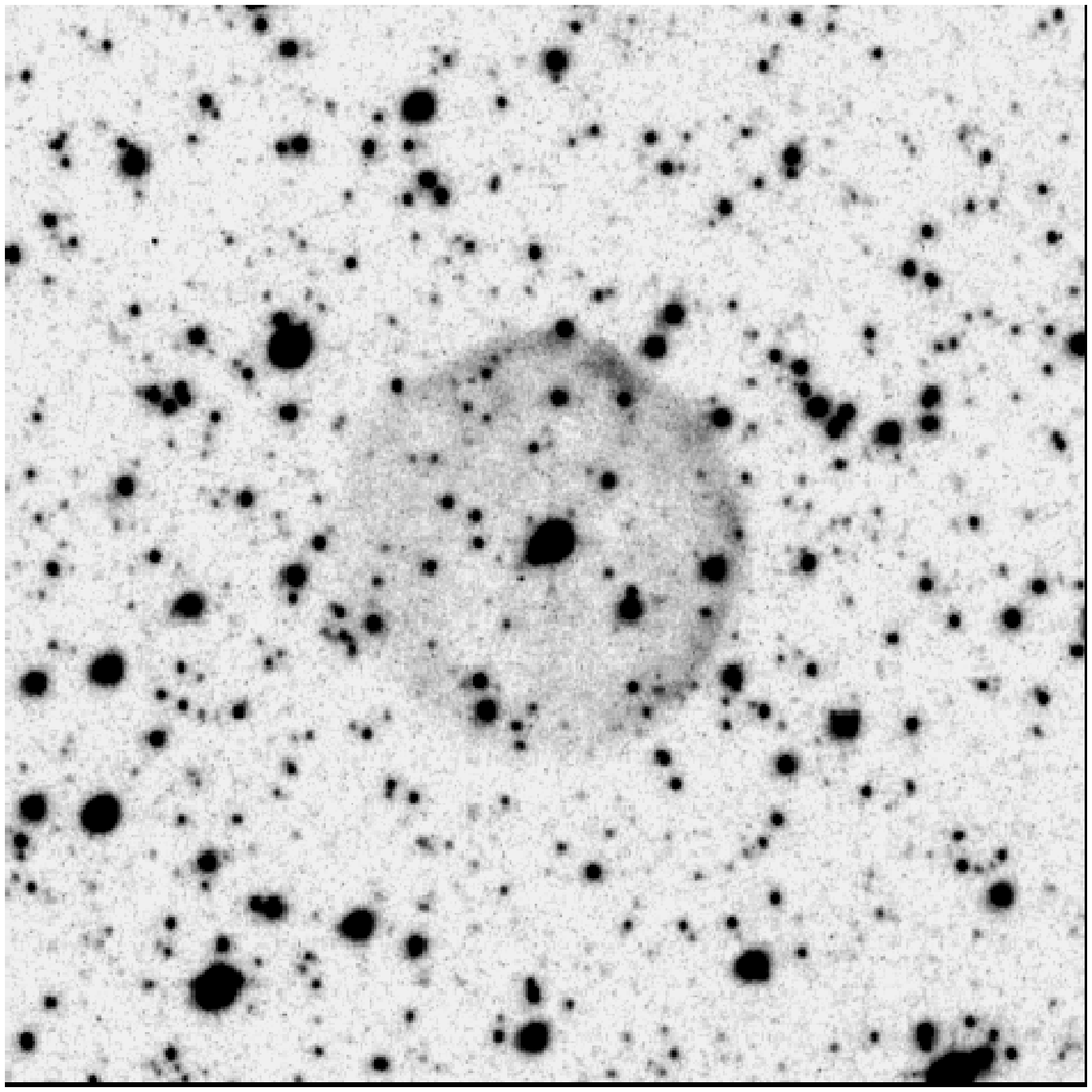}
\caption{Images of various PN mimics, adapted from DSS and SHS images.  Top row, from left to right: (1): NGC 6164/65, a bipolar ejecta nebula around the Of star HD~148937 ($B_{J}$ image, 10\arcmin\ wide). (2): The faint shell around the Wolf-Rayet star WR~16 (\ha, 10\arcmin).  (3): The WR ejecta nebula, PCG~11 (\ha, 5\arcmin).  (4) Longmore 14, a reflection nebula  ($B_{J}$, 5\arcmin). 
Second row:  (1):  Abell~77 (Sh~2-128), a compact HII region ($R_{F}$, 5\arcmin).  (3):  The low-surface brightness, diffuse HII region, vBe 1 (\ha, 10\arcmin). (3):  Sh~2-174, a \HII\ region ionized by a hot white dwarf ($R_{F}$, 20\arcmin).  (4):  Bipolar symbiotic outflow, He~2-104 (\ha, 3\arcmin).  
Bottom row:  (1): Faint bipolar nebula around the B[e] star, He 3-1191 (\ha, 4\arcmin).   (2):  The old nova shell around GK~Per ($R_{F}$, 5\arcmin).  (3):  Blue compact galaxy He~2-10 ($R_{F}$, 5\arcmin).  (4):  The true PN PHR~J1424-5138 as a comparison object.  Note the unusually bright CS relative to the low nebular surface brightness for this particular PN (\ha, 5\arcmin).}
\label{image_collage}
\end{center}
\end{figure*}

The similarity in the morphologies of some bipolar PN like NGC~6537 and Hubble~5 with the symbiotic outflow He~2-104 (see Figure~\ref{image_collage}, and \S\ref{sec:symbio}) is illustrative.\footnote{For detailed images of these and other objects, see http://www.astro.washington.edu/users/balick/PNIC/}  It is clear that the potential for such contaminants to appear in the various catalogues of PN is high.  While many PN have a canonical morphology and spectrum (see Figure~\ref{spectrum_collage}), we have found that a multiwavelength approach is the only way to reject unusual candidates unequivocally (Parker et al. 2006, Miszalski et al. 2008, and Cohen et al. 2007, 2010).  Therefore it is germane to investigate PN mimics in detail (see Acker et al. 1987; Kohoutek 2001; Parker et al. 2006), so a much needed review is given in this section.

\subsection{HII regions and Str\"omgren zones}  
\begin{figure*}
\begin{center}
\includegraphics[scale=0.3,angle=-90]{NGC7009_spec.ps}
\includegraphics[scale=0.3,angle=-90]{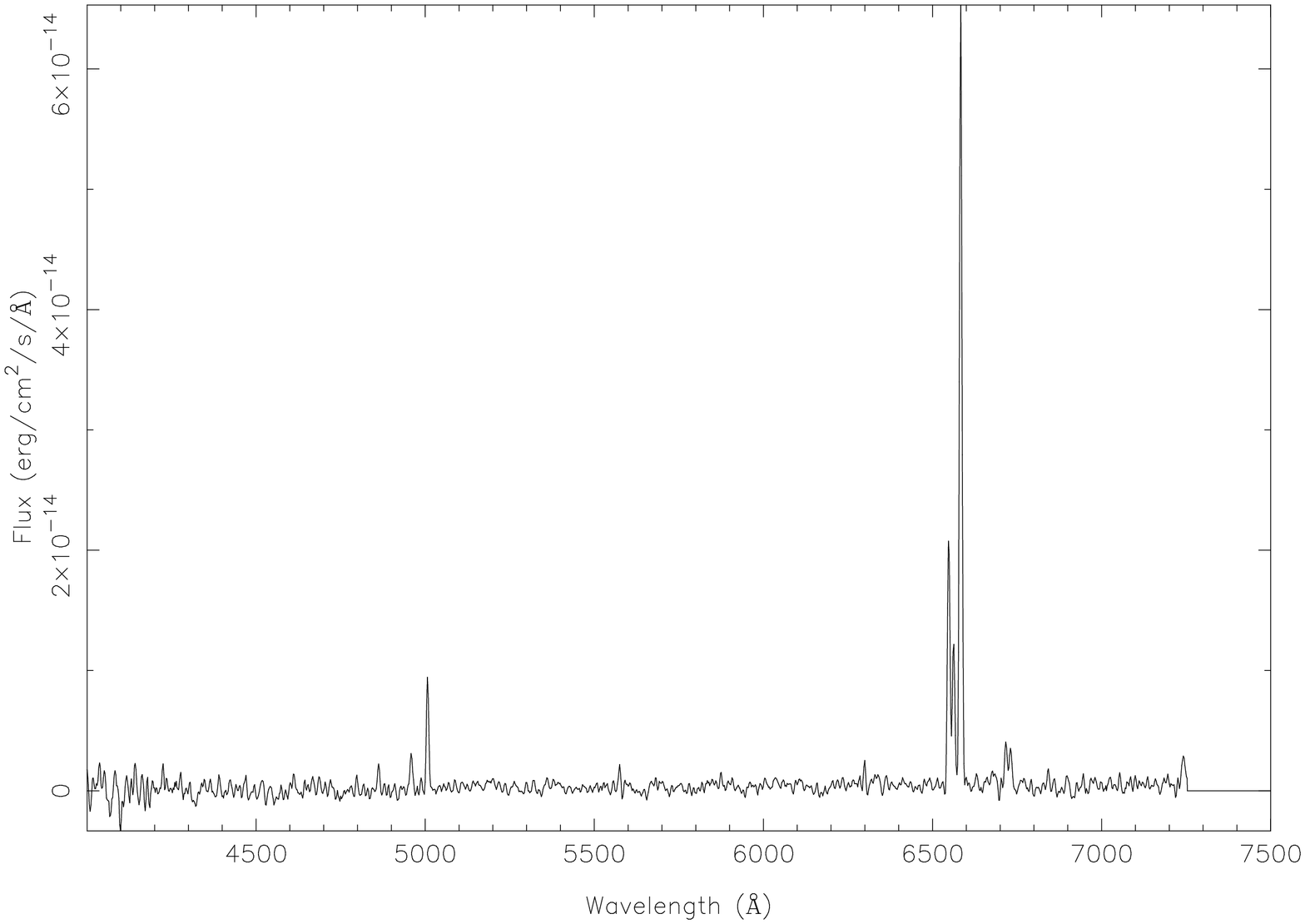}
\includegraphics[scale=0.3,angle=-90]{RPZM31_spec.ps}
\includegraphics[scale=0.3,angle=-90]{PFP1_smooth.ps}
\includegraphics[scale=0.3,angle=-90]{K1-27_smooth.ps}
\includegraphics[scale=0.3,angle=-90]{PMR3_spec.ps}
\caption{Representative PN spectra showing the diversity in line-ratios amongst the class.  Top row: (left) NGC 7009, a typical PN of medium-high excitation; (right) RCW~69 (Frew, Parker \& Russeil 2006), a reddened, bipolar Type~I PN with very strong \NII\ lines;
Second row: (L) RPZM~31, a compact VLE PN with no \OIII\ emission;  (R) PFP~1 (Pierce et al. 2004), a highly-evolved, non-Type I PN with quite strong \NII\ lines; 
Bottom row: (L) K~1-27, a peculiar PN with very high excitation --- note the very strong $\lambda$4686 \HeII\ line and the complete absence of \NII\ and \SII\ lines; (R) PHR J1641-5302, a high excitation PN around a [WO4] central star; note the broad emission feature from the CS near 4650\AA\ due to CIII and HeII, and the CIV feature at 5806\AA.  This PN also has a dense unresolved core (Parker \& Morgan 2003) with strong $\lambda$4363 emission.   All spectra cover the same wavelength range (4000--7500\AA) and were taken with the SAAO 1.9-m reflector, except for RPZM~31 which was observed with 6dF on the 1.2-m UKST; some bright lines have been truncated for clarity.}
\label{spectrum_collage}
\end{center}
\end{figure*}

\begin{figure*}
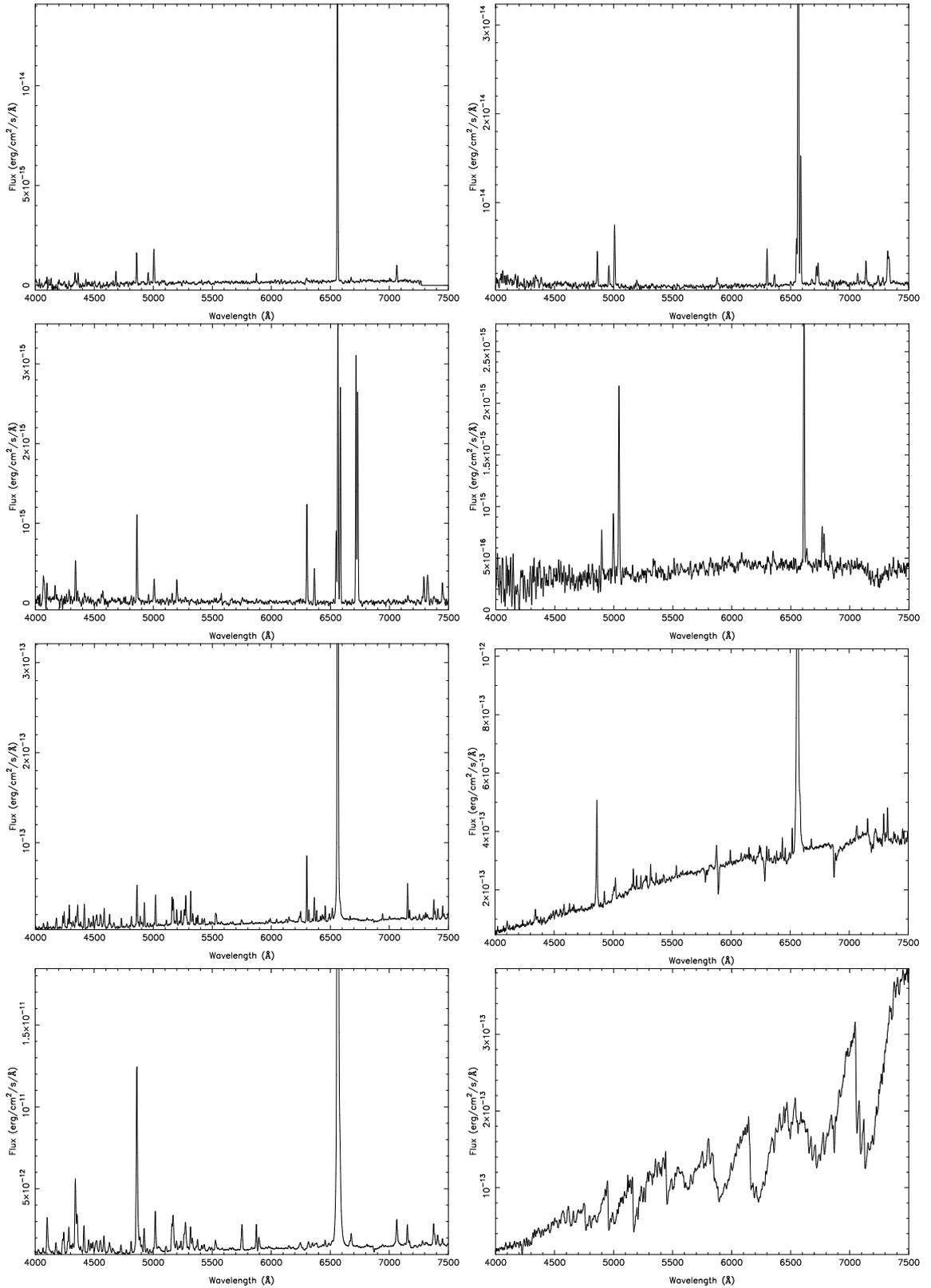

\begin{center}
\includegraphics[scale=0.3,angle=-90]{PTB15_scaled.ps}
\includegraphics[scale=0.3,angle=-90]{M2-62_smooth.ps}
\includegraphics[scale=0.3,angle=-90]{PHR0818-4728.ps}
\includegraphics[scale=0.3,angle=-90]{PHR0950-5223_spec.ps}
\includegraphics[scale=0.3,angle=-90]{He3-1191_spec.ps}
\includegraphics[scale=0.3,angle=-90]{BICru_spec.ps}
\includegraphics[scale=0.3,angle=-90]{eta_car_spec.ps}
\includegraphics[scale=0.3,angle=-90]{PHR1649-4701_spec.ps}
\caption{Representative spectra of a range of PN mimics. Top row: (left)  PTB~15, a ring-shaped nebula with a CS that shows symbiotic characteristics, i.e. $\lambda$4363 $\simeq$ \hg;  (right) M~2-62, a reddened, compact HII region;
Second row: (L) PHR~J0818-4728, a compact knot in the Vela supernova remnant, showing strong \SII\ and \OI\ lines;  (R) PHR~J0950-5223, a emission-line galaxy with {\it cz} = 2300 \kms; 
Third row: (L) He~3-1191, a B[e] star surrounded by a faint bipolar outflow --- note the intense, broad \ha\ line, strong \OI$\lambda$6300, and the forest of {Fe\,\textsc{ii}} and {[Fe\,\textsc{ii}]} lines;  (R) BI Crucis, a low-excitation symbiotic star with weak B[e] characteristics; 
Bottom row: (L) Eta Carinae, a LBV star with a rich emission-line spectrum which resembles a B[e] star, but there is no sign of any \OI$\lambda$6300 emission;  (R) An M-type giant where the continuum peak near 6550\AA\ can mimic \ha\ emission in narrowband filters.   All spectra cover the same wavelength range (4000--7500\AA) and were taken with the SAAO 1.9-m reflector.}
\label{spectrum_collage2}
\end{center}
\end{figure*}

 A number of compact \HII\ regions have been picked up by the numerous objective-prism surveys over the years, with several being misidentified as PN (see  figure~\ref{spectrum_collage}).  Other \HII\ regions have symmetrical forms which have been confused with PN.  Likely examples include He~2-77 and He~2-146 (Cohen et al. 2010), We\,1-12 (Kaler \& Feibelman 1985; Kimeswenger 1998), Abell\,77 (Bohigas \& Tapia 2003; Figure~\ref{image_collage}), and M~2-62 (Figure~\ref{spectrum_collage2}).  The peculiar nebula M~1-78 (e.g. Gussie 1995) is now interpreted as a compact \HII\ region contaminated by N-rich ejecta from a WR star (Martin-Hern\'andez et al. 2008).

NGC 2579 is another interesting case: this distant \HII\ region has also been classified as a planetary nebula (=Ns 238; Nordstr\"om 1975; Schwarz, Corradi \& Melnick 1992), a cometary nebula (Parsamyan \& Petrosyan 1979) and an ordinary reflection nebula in the past (see the discussion by Copetti et al. 2007).  

In general, consideration of the nebular spectral features, radio fluxes, environment, and near- and mid-IR characteristics (including any embedded sources) is usually enough to differentiate between (ultra-) compact \HII\ regions and PN.  In these cases, the optical nebula is typically only the ionized component of a much larger mid-IR structure.  Further details are given in Cohen \& Parker (2003), Parker et al. (2006) and Cohen et al. (2010).  Compact \HII\ regions in external galaxies can also be confused with PN, but generally appear more extended if not too distant, usually have lower ionization emission lines, and often show strong continua from embedded OB stars (Jacoby 2006). 

Not all \HII\ regions surround massive stars, a fact that is often not fully appreciated.  Recent work by our team has shown that several nearby nebulae currently accepted as PN are simply Str\"omgren zones in the ISM (Frew \& Parker 2006; Madsen et al. 2006; Frew et al. 2010, in preparation), each one ionized by a hot white dwarf or subdwarf star --- the original PN, if any, having long since dissipated.  That is, these nebulae are low-mass \HII\ regions.  They may have emission-line ratios which mimic true PN as the ionizing stars can still be very hot, e.g. Hewett~1 (Chu et al. 2004) and Sh\,2-174 (Madsen et al. 2006; Figure~\ref{image_collage}).


\subsection{Reflection nebulae}  
A few symmetrical reflection nebulae have been misclassified as PN over the years, such as NGC~1985 (Sabbadin \& Hamzaoglu 1981; Lutz \& Kaler 1983) and Lo~14 (Longmore 1977; Figure~\ref{image_collage}).   Furthermore, the occasional `cometary' reflection nebula (e.g. Parsamyan \& Petrosyan 1979; Neckel \& Staude 1984) has been confused with a PN on morphological grounds. In general, the optical and near-IR broadband colours and continuous spectra of reflection nebulae allow them to be readily differentiated from PN.  Note that from the initial visual search of the AAO/UKST \ha\ Survey, reflection nebulae were weeded out from PN candidates by comparing short-red and $I$-band images with H$\alpha$ images. 

\subsection{Population~I circumstellar nebulae}
Circumstellar nebulae are found around OB type stars, luminous blue variables (LBVs), Population~I WR stars, and supergiant B[e] stars (see {\S\ref{sec:Be_stars}).  The peculiar nitrogen-rich, bipolar nebula NGC\,6164--65 (Figure~\ref{image_collage}) surrounding the Ofp star HD~148937, was initially catalogued as a PN by Henize (1967).  Another example is the annular nebula He~2-58 (Thackeray 1950; Henize 1967), associated with the LBV star AG\,Carinae.   LBVs are variable, early-type hypergiants which display rich emission-line spectra from H, He and a number of permitted and forbidden metal species.  These lines help to differentiate LBVs from B[e] stars, other emission-line stars, and compact PN (though see Pereira et al. 2008, for a discussion of the problems in classification).  

Another example of an LBV nebula is the famous Homunculus (e.g. Smith 2009) ejected by the eruptive variable $\eta$\,Carinae.  This presents a morphology that closely resembles a bipolar PN (e.g. Hb~5), but is one to two orders of magnitude more massive.  Remarkably, it was formed over a very brief period ($<$5~yr; Smith 2006) during the Great Eruption of this star observed in the 1840s (Frew 2004).  Several other candidate LBVs have surrounding nebulae that are morphologically similar to PN: examples are Hen~3-519 (Stahl 1987), Wray~17-96 (Egan et al. 2002), and HD\,168625 (Smith 2007). 

Circumstellar nebulae are often found around WR stars, which are evolved, massive stars with exceptionally strong stellar winds that shape the surrounding ISM.  Marston (1997) found that $\sim$35\% of 114 southern WR stars have detectable circumstellar nebulae in \ha.  We note that a few WR ejecta nebulae have been confused with PN in the past, e.g. M~1-67 (Merrill's star), which was suspected to be an unusual PN for many years (e.g. Perek \& Kohoutek 1967), while PCG~11 (Cohen, Parker \& Green 2005; Figure~\ref{image_collage}) is another example showing a PN-like morphology.  Other WR nebulae are simple wind-blown shells in the ISM; examples are NGC~6888 (the Crescent Nebula) and the near spherical limb-brightened shell around WR~16 (Marston et al. 1994; Figure~\ref{image_collage}).

\subsection{Supernova remnants (SNRs)}\label{sec:SNR} 
At least one filamentary nebula initially classed as a PN has turned out to be a SNR, e.g. CTB~1 = Abell~85 (Abell 1966).  Conversely, the one-sided filamentary PN, Abell\,21 and Sh~2-188, have been misclassified as SNRs in the past.   A few individual compact knots of extended SNRs have also been misidentified as PN candidates, such as H\,2-12 which is a bright knot in Kepler's SNR (Acker et al. 1987), and PHR\,0818-4728, an isolated knot in the Vela SNR (Figure~\ref{spectrum_collage2}).  Attention is also drawn to the unusual filamentary nebula FP~0821-2755.  It has been classified in the literature as a possible SNR by Weinberger et al. (1998) and Zanin \& Kerber (2000), though it is more likely to be a peculiar Type~I PN instead (Parker et al. 2006).   In general there can sometimes be confusion between old arcuate PN and partially fragmented, old SNRs (see Stupar, Parker, \&  Filipovi\'c 2008), especially if the emission-line ratios are equivocal.

SNRs in external galaxies may possibly be confused with PN (Ciardullo, Jacoby \& Harris 1991; Jacoby 2006), but are generally more extended (if not too distant to be resolved), and like their Galactic counterparts, can be generally separated from PN using optical emission-line diagnostic diagrams (see \S\ref{sec:diagnostic}).

\subsection{Herbig-Haro (HH) objects and young-stellar objects (YSOs)}  
For detailed reviews, including a discussion of the spectra of HH objects, see Dopita (1978), Raga, B\"ohm \& Cant\'o (1996) and Reipurth \& Bally (2001).  In most cases, HH objects can usually be readily differentiated from PN, due to their distinctive morphology, the presence of shock-excited spectral features such as strong {[\rm S \sc ii]}, {[\rm N\sc i]} and {[\rm O\sc i]} emission (e.g. Cant\'o 1981, Cappellaro et al. 1994), and their proximity to star-forming regions and associated molecular clouds with high levels of surrounding extinction.   Due to the combination of low excitation and high density, the blue $\lambda\lambda$4068,4076\AA\ [S\,{\sc ii}] lines are often particularly strong relative to H$\beta$ in HH objects, if the reddening is not too severe.  We note that two of the four compact PN candidates identified by Viironen (2009a) are located in areas of very heavy obscuration.  The blue [S\,{\sc ii}] lines are very strong relative to H$\beta$ in both candidates, though these authors pointed out that some rare, shock-excited pre-PN appear to have similar spectral features (e.g. Goodrich 1991), so the jury remains out on their nature.  

Related to HH objects are the T Tauri stars and Herbig Ae/Be stars.  Many HAeBe stars have rather similar spectra to the B[e] stars but often have reversed P Cygni profiles, a signature of accretion or matter infall (Lamers et al. 1998), and are closely associated with regions of active star formation, which can be used to ascertain their true character.

\subsection{B[e] stars}\label{sec:Be_stars}
Be stars are B-type main-sequence, subgiant or giant stars with prominent Balmer emission (see Porter \& Rivinius 2003 for a review), while the B[e] stars have additional forbidden emission lines of species such as  {[\rm Fe \sc ii]}, {[\rm O \sc i]}  and {[\rm O \sc ii]}, as well as a strong IR excess.  Several B[e] stars were confused with compact or unresolved PN on early objective-prism plates, e.g. M~1-76 (Sabbadin \& Bianchini 1979).  The B[e] stars are not a homogenous class of objects, and are best referred to as a collective of stars showing the `B[e] phenomenon'  (Lamers et al. 1998; Miroshnichenko 2006).  The group includes several evolutionary phases that are, in the main, unrelated to each other.   Lamers et al. (1998) subdivided them into (a) sgB[e], supergiants, (b) pre-main sequence HAeB[e]stars, (c) cPNB[e], compact planetary nebula central stars, (d) SymB[e], symbiotic B[e] stars, and (e) unclassified stars, denoted unclB[e].   Most stars in this latter group are now placed in the FS~CMa class (Miroshnichenko 2007).

An important subclass in the context of this review are the cPNB[e] stars, many of which are associated with highly collimated, bipolar outflows.  Examples are M~2-9 (Schwarz et al. 1997; Phillips \& Cuesta 1999; Livio \& Soker 2001), Menzel~3 (Smith 2003; Smith \& Gehrz 2005), He~3-401 (Garc\'ia-Lario, Riera \& Mampaso 1999; Hrivnak et al. 2008; Pereira et al. 2008), He~3-1475 (Riera et al. 1995) and He~2-90 (Guerrero et al. 2001; Kraus et al. 2005).  Another is He~3-1191 (Figures~\ref{image_collage} \& \ref{spectrum_collage2}), a B[e] star with a dusty circumstellar disk (Lachaume et al. 2007); its evolutionary state is unknown.  

Many of these nebulae are simply called proto- or pre-PN\footnote{Sahai et al. (2007) have argued for adoption of the term {\sl pre-PN} to refer to these objects.} in the literature, but there are enough morphological and spectroscopic similarities amongst these objects (e.g. strong Balmer emission plus a rich zoo of permitted and forbidden metal lines, along with a strong IR excess and extremely collimated outflows), that we can suggest they might form a homogenous class.  The evolutionary state of the unusual B[e] star MWC~922 is uncertain, but it too has a surrounding strongly bipolar nebula, nicknamed the Red Square (Tuthill \& Lloyd 2007).  Other nebulae, like the pair of bipolar outflows or ``wing-nuts'', He~2-25 and Th~2-B (Corradi 1995) have central stars with strong $\lambda$4363 \OIII\ emission, characteristic of symbiotic outflows (see Section~\ref{sec:symbio});  M~1-91 is another closely similar nebula (Rodr\'iguez, Corradi \& Mampaso 2001). 

Should these B[e] outflows be classed as pre-PN, as has been suggested for a number of their class?  Gathering up the scanty distance information from the literature suggests not.  For example, Sahai et al. (2007) estimated distances and angular sizes for 23 pre-PN.  Using this data, the median diameter of the sample is found to be 0.05\,pc (recall \S\ref{sec:PN_definition}).  In contrast,  the Ant nebula, Mz~3, is also called a bipolar pre-PN.  However, published distances range from 1.8--3.3\,kpc (Cohen et al. 1978; Van der Veen, Habing \& Geballe 1989; Kingsburgh \& English 1992; Smith 2003), which leads to a nebular major axis length of 0.4--0.8\,pc, or more than an order of magnitude larger than most bona fide pre-PN.  There is also evidence of a cool giant (Smith 2003) in this system, so it may be best classed as a symbiotic B[e] star.  Another pre-PN candidate is OH\,231.8+4.2 (the Calabash nebula), which has a well determined phase-lag distance of 1.3 kpc (e.g. Kastner et al. 1992).  At this distance, the major axis of the nebula subtends 0.4 pc.  Combined with the presence of a heavily obscured Mira star, this indicates the Calabash is probably a symbiotic outflow (see \S~\ref{sec:symbio}) and should be removed from the pre-PN class. 

Observationally, the sgB[e] stars appear to have fast rotation and/or the presence of an accretion disk, which may be hinting at a binary origin, though other mechanisms have been proposed (Polcaro 2006).  Several of the most luminous sgB[e] stars appear to have large, often bipolar nebular shells associated (e.g. Marston \& McCollum 2008; cf.  Smith, Bally \& Walawender 2007).  In addition, Podsiadlowski, Morris \& Ivanova (2006) propose that the merger rate of massive stars is compatible with the B[e] supergiant formation rate, suggesting that sgB[e] stars are primarily a merger product.  Kraus et al. (2005) found that the central star of the PN-like nebula He~2-90 is also rotating very rapidly.  Is this object also a merger product, or at least the spawn of a strong mass exchange process, as has been postulated for the FS~CMa stars (Miroshnichenko 2007)?  It seems logical to propose that a binary channel is applicable to all stars showing the B[e] phenomenon, regardless of evolutionary state, but obviously further work is needed (Kraus et al. 2009).

\subsection{Symbiotic Stars and Outflows}\label{sec:symbio} 
A wide range of stars with emission lines have been discovered over the years from low angular and spectral resolution objective-prism surveys.  As a result, many have been misinterpreted as compact and `stellar' PN in previous catalogues (see Acker et al. 1987).  The most common stellar-appearing PN mimics are the symbiotic stars.  The traditional definition encompasses those interacting binary objects that show a composite spectrum with emission lines of H\,I and He\,I (and often {[\rm O \sc iii]} and He\,II) present in conjunction with an absorption spectrum of a late-type giant star (Allen 1984; Kenyon 1986; Belczy\'nski et al. 2000).   In reality, they show a wide variety of spectral characteristics; see Munari \& Zwitter (2002) for a comprehensive spectral atlas of symbiotic stars.

Symbiotic stars are subdivided into S-type systems which contain a normal (\emph{stellar}) red giant which dominates the near-IR colours, and D-type systems where the central binary contains a Mira and the near-IR colours are dominated by heated \emph{dust}.  An additional class, the D$^{\prime}$ symbiotics, contain F-, G- or K-type giants, and contain cooler dust than the D-types (Allen 1982).  

A class of \emph{resolved symbiotic outflows} around D-type symbiotic Miras (Corradi et al. 1999; Corradi 2003) often show close morphological similarity to highly collimated bipolar PN.  Even though some authors use the term `symbiotic PN', arguing that symbiotics and PN (and pre-PN) have physical links, it is likely that most are not true PN (Corradi \& Schwarz 1995; Corradi 1995; Kwok 2003, 2010; cf. Nussbaumer 1996).  

We use the definition given by Corradi (2003)  which states that in resolved symbiotic nebulae, the ionized gas originates from a star in the RGB or AGB phase.  Corradi (2003) has also used a simple statistical argument to show it is highly improbable that a true PN central star would have a binary Mira companion.  Nevertheless, evolutionary links between the symbiotic outflows and PN are likely: many current symbiotic systems may have been PN in the past when the WD companion was at an earlier phase in its evolution.  Similarly, many PN with binary central stars may go through a symbiotic phase in the future, when the companion star evolves to the AGB. 

For an illustration of the confusion present in the literature, the bipolar outflow around BI~Crucis has always been considered to be a symbiotic nebula since discovery (Schwarz \& Corradi 1992; Corradi \& Schwarz 1993), as has the bipolar nebula around the nearby Mira star R~Aquarii (Kaler 1981 and references therein).  However, He\,2-104, the Southern Crab (Lutz et al. 1989; Corradi \& Schwarz 1993; Corradi et al. 2001; Santander-Garc\'ia et al. 2008) while initially classified as a PN (Henize 1967), is ostensibly of the same class --- all three nebulae have Mira stars as their central engines.  

Another unusual object, RP\,916, a recently discovered PN candidate in the LMC (Reid \& Parker 2006b), is shown by Shaw et al. (2007) to be a very unusual nebula.  However, the very red IR colours of the central source, its intrinsic variability, along with the extreme bipolar morphology, may indicate that it is a symbiotic outflow analogous to its better studied Milky Way cousins.  Its large size of $\sim$0.75\,pc is notable, but not unprecedented amongst symbiotic nebulae (Corradi 2003).  In the future, near-IR spectroscopic observations should shed light on the true nature of this unusual object.  

Some of the most problematic objects belong to the yellow D$^{\prime}$-type symbiotic class (Jorrisen 2003; Frankowski \& Jorissen 2007).  Several have been catalogued as `stellar' PN in the past, such as M\,1-2 (Siviero et al. 2007) and Cn\,1-1 = HDE\,330036 (Lutz 1984; Pereira, Smith \& Cunha 2005), and possibly PC\,11 (Guti\'errez-Moreno \& Moreno 1998; Munari \& Zwitter 2002) and PM~1-322 (Pereira \& Miranda 2005).  On the other hand, some D$^{\prime}$-type symbiotic stars have resolved, low-surface brightness, slowly expanding nebulae around them which appear to be fossil PN (Corradi 2003; Jorissen et al. 2005; cf. Schmid \& Nussbaumer 1993).  Probable examples include AS~201 (Schwarz 1991) and V417~Cen (Van Winckel et al. 1994;  Corradi \& Schwarz 1997).

\subsubsection{EGB~6 and its kin}
We give an interlude at this point to discuss the nearby ($\sim$500\,pc) low-surface brightness, bona fide PN, EGB~6 (Ellis, Grayson \& Bond 1984), as it may have features in common with objects like AS~201.  The central star of EGB~6 contains an unresolved, high-density, emission-line nebula (Liebert et al. 1989), and as such has observational similarities with the symbiotic stars.  Bond (2009) summarises the properties of the compact nebula; $HST$ imaging shows it is actually associated with a resolved companion to the true CS.  However, the origin of the compact nebula is unclear; Bond (2009) speculates that it could arise in an accretion disk of material around the companion captured from the outflow that produced the surrounding large PN (see also Bilikova et al. 2009).  

Recent observations show that EGB~6 is no longer unique as other PN with unresolved inner nebulae are now known, including NGC~6804 (A. Peyaud et al. 2010, in preparation), Bran~229 (Frew et al. 2010, in preparation), and M~2-29 (Torres-Peimbert et al. 1997; Hajduk, Zijlstra \& Gesicki 2008; Miszalski et al. 2009a; Gesicki et al. 2010), all of which have evidence for a cool companion to the CS.   EGB~6 can be thought of as a prototype of a class of otherwise normal PN holding compact, high-density, emission nebulae around a central binary star (see Gesicki et al. 2010, for a further discussion).

Parker \& Morgan (2003) found that PHR~J1641-5302, a high excitation PN with a [WO4] CS, also has a  unresolved high-density nebula around the CS as revealed by strong $\lambda$4363 emission (Figure~\ref{spectrum_collage}).  We also note PTB~15 (Boumis et al. 2003) which appears as a thick annulus with an obvious off-centre CS on SHS \ha\ images.   Note this object was originally discovered as part of the MASH survey and designated PHR\,1757-1711.  It was observed spectroscopically in July 2000 (see Figure~\ref{spectrum_collage2}) where the unusual spectrum of the central star later led to it being removed from the final MASH catalogue (cf.  Parker et al. 2003).  Our spectrum across the CS shows strong \OIII$\lambda$4363 emission superposed on the nebular background.  It is currently unclear if this is a pole-on symbiotic outflow or an EGB~6-like PN.  The 2MASS colours are very red, but the luminosity at K$_{s}$ is less than expected for an AGB or RGB star for our preferred distance range of 3--7 kpc.  We plan further observations of this interesting nebula.  

In a similar vein to both PN and B[e] stars, we consider that the `symbiotic phenomenon' is quite heterogeneous and is manifested in a wider range of evolutionary states than is usually appreciated.  This leads to potential confusion between some symbiotic stars and emission-line cores in PN.  Another example is WhMe~1 = IRAS~19127+1717 (Whitelock \& Menzies 1986); this has been thought to be either a very compact PN or a symbiotic star (De Marco 2009).  The spectrum has the typical emission lines of a symbiotic system, but the underlying continuum is that of a B9~III--V star (Whitelock \& Menzies 1986).  We further note that it is completely stellar on all available images, so is it related to the EGB~6-like PN, or is it a genuine yellow symbiotic?  An argument against the latter is that the B9 star is expected to have a negligible wind, so it is not expected to be the mass donor in a classical symbiotic system; there is also no sign of any CV characteristics.  If instead the B9 star has accreted a compact disk of material during a recent PN phase of the companion star, this would explain the observed spectral features.  A deep search for a fossil PN around this system might be rewarding.  

We note two other evolved PN which have interesting emission-line nuclei, both showing the Ca\,II infrared triplet in emission; these are He~2-428 (Rodr\'iguez et al. 2001) and IPHAS~PN-1 (Mampaso et al. 2006), though both appear to be beyond 3\,kpc in distance, suggesting similar nuclei are rare.  Ca\,II emission is commonly seen in B[e] stars, some pre-PN, symbiotic stars, cataclysmic variables, YSOs and T Tauri stars (e.g. Persson 1988).  Both He~2-428 and IPHAS~PN-1 are strongly bipolar nebulae with unresolved, very dense cores ($n_{e}$  $>$10$^{10}$\,cm$^{-3}$)  (Mampaso et al. 2006).  Neither star shows the optical \FeII\ lines typical of the B[e] stars because these lines are collisionally quenched at very high densities; there may be no lower-density zones in the cores of these `PN'.   We emphasise the surprising diversity in the spectra of emission-line CS of PN-like nebulae, and more work is need to disentangle the relationships between the various sub-classes, and between these objects and the B[e] and symbiotic stars.

\subsection{Late-type stars} 

On small-scale objective-prism red plates, the spectra of ordinary M-type giant stars can present apparent `emission' near H$\alpha$ due to the relative brightness of the red continuum around 6600\AA\ compared to the deep absorption dips from TiO band-heads at adjacent wavelengths. These bands continue to rise strongly redward but they cannot be used to aid in discrimination due to the emulsion cut-off at $\sim$7000\AA\ that corresponds to a band dip.  

More recently, many compact PN candidates have been discovered from the H$\alpha-R$ versus $R$ plots obtained from SuperCOSMOS IAM data (Parker et al. 2005) and from SHS quotient imaging (Miszalski et al. 2008).  Although many sources with an apparent excess are true \ha\ emitters, the use of $I$-band imagery enables the M-type stellar contaminants to be effectively removed from lists of PN candidates when the excess is less extreme.  Of course follow-up spectroscopy settles the matter in uncertain cases.  The northern IPHAS survey uses a combination of narrowband \ha, and Sloan $r${$^\prime$} and $i${$^\prime$} filters, so ordinary late-type stars are not a contaminant in this survey (Corradi et al. 2008).  

Mira variables can also have the hydrogen Balmer lines in emission.  The emission lines are caused by pulsation-induced shock waves in the Mira's atmosphere, and usually more prominent near maximum light.   In addition, variable stars can be sometimes misidentified as compact PN candidates in any techniques which use on-band/off-band image subtraction or blinking.  This can occur if the exposures are not contemporaneous and the star has changed brightness between exposures (Jacoby 2006). 

\subsection{Cataclysmic variable stars (CVs)}\label{sec:CVs}
This large class of variable stars includes classical and recurrent novae (Bode 2009), nova-like variables and dwarf novae.  CVs are semi-detached binaries in which a WD star is accreting material from a close, Roche lobe-filling companion, usually a low-mass, main sequence star.

Erupting novae in the nebular phase show strong Balmer and [O~{\sc iii}] emission and were sometimes confused in the past with unresolved PN on objective-prism plates.  There are even a few old novae in the NGC/IC catalogues (e.g. IC\,4544 = IL~Normae and IC\,4816 = V1059~Sgr), erroneously classified as `gaseous nebulae'.  Contaminating novae have also been found in recent CCD surveys for PN in the Magellanic Clouds (Jacoby 2006).  

At later stages in the evolution of classical and recurrent novae, resolved shells become visible, comprised of gaseous ejecta produced in the nova eruption.  Nova shells show both spherical and axisymmetric structures (e.g. Downes \& Duerbeck 2000), and can have spectral signatures reminiscent of PN.  Typically about 10$^{-4}$\,$M_{\odot}$ of ejecta is produced (Cohen \& Rosenthal 1983), or two to four orders of magnitude less mass than is contained in a PN shell.  Recently, Shara et al. (2007) reported a large extended fossil nova shell around the dwarf nova, Z Cam.  The shell around GK Persei (Nova Persei 1901) is the best-known example of the class (Figure~\ref{image_collage}) but is a rather atypical remnant in a number of ways (Bode 2009).  GK Persei is of further interest due to the presence of a possible ancient PN surrounding it (see Bode et al. 1987; Bode 2004).  Another example of a likely old PN around a classical nova, V458~Vul, has been recently discovered by Wesson et al. (2008b).

\subsection{Miscellaneous emission nebulae}\label{sec:miscell} 
This is a rather heterogeneous class of objects.  Emission nebulae with `PN-like' optical spectra and morphologies (Remillard, Rappaport \& Macri 1995) are known to be associated with luminous supersoft X-ray sources (SSS), which are likely to be WDs in binary systems accreting matter at a high enough rate to allow quasi-steady burning of hydrogen in their envelopes (Kahabka \& van den Heuvel 1997).   

Pulsar wind nebulae can also have similar optical spectra.  For example, the compact core of the supernova remnant CTB~80 is morphologically like a PN in {[\rm O \sc iii]} light (Blair et al. 1984), has a PN-like optical spectrum, but has very strong {\sl non-thermal} radio emission.  It is probably a pulsar wind bowshock seen relatively head on (Hester \& Kulkarni 1988). 

There are also two optically-emitting bowshock nebulae known to be associated with nova-like CVs, and both were, at least initially, considered as PN candidates.  They are EGB~4 associated with BZ~Cam (Ellis et al. 1984; Greiner et al. 2001) and Fr~2-11, associated with a previously unnoticed nova-like variable, V341~Ara (Frew, Madsen \& Parker 2006; Frew et al. 2010, in preparation).

Furthermore, the peculiar PN-like nebula Abell~35 (Jacoby 1981; Hollis et al. 1996) may be a bowshock nebula inside a photoionized Str\"omgren sphere in the ambient ISM (see Frew 2008 for a discussion).  Indeed, the morphologies and optical spectra of EGB~4, Fr~2-11, and Abell~35 are all quite similar hinting that the emission is due to a combination of shock excitation and photoionization.   Poorly known PN candidates in the current catalogues (e.g. BV~5-2, PHR~J1654-4143) may prove to be of this type after a detailed investigation of their ionizing stars and surrounding nebulae.

\subsection{Galaxies} 
Low-surface brightness (LSB) dwarf galaxies have been confused with evolved PN in the past on morphological criteria alone.  Deep imaging (occasionally resolving the stellar population; e.g. Hoessel, Saha \& Danielson 1988) or spectroscopy, which detects the redshift, has confirmed their identity as galaxies.  Conversely, a few candidate LSB galaxies, especially those closer to the Galactic plane, have been shown to be bona fide PN (e.g. Hodge, Zucker \& Grebel 2000; Makarov, Karachentsev \& Burenkov 2003).

A few emission-line galaxies appeared in the early PN catalogues (e.g. Acker, Stenholm \& V\'eron 1991).  Several galaxy sub-types have strong Balmer and forbidden emission lines, and are potentially confused with PN on objective prism plates, where the lack of spectral resolution may preclude an extragalactic identification for low redshift examples.   He\,2-10 (Figure~\ref{image_collage}) and IC~4662, classed as possible PN by Henize (1967), are now known to be nearby starburst dwarf galaxies.  Additionally, there were five emission-line galaxies in the preliminary MASH list until confirmatory spectroscopy revealed their redshifts, e.g. PHR~J0950-5223 (Figure~\ref{spectrum_collage2}).

Lastly, distant emission-line galaxies can contaminate on-band/off-band [O\,{\sc iii}] surveys for PN in galaxies in and beyond the Local Group (Jacoby 2006).   Ly-$\alpha$ emission from starbust galaxies at redshift, {\it z} = 3.1 falls in the bandpass of [O\,{\sc iii}] $\lambda$5007 interference filters, as do [O\,{\sc ii}] emitters at {\it z} = 0.34.  Rarely, closer emission-line galaxies at {\it z} = 0.03 can have \hb\ shifted into the filter bandpass (e.g. Jacoby 2006), but these can generally be eliminated on other grounds.

\subsection{Plate defects and flaws}
Many examples of emulsion and processing defects are present on Schmidt plates and films (see Parker et al. 2006, for additional details).  Comparison of first and second epoch plates can generally eliminate these flaws easily.  However, when only a single epoch plate is available, the true identity of a PN candidate is less obvious. A number of entries in the Abell (1966) catalogue were only visible on the POSS red plate, and some of these have turned out to be flaws (e.g. Abell~17 and Abell~32; K. Wallace, 2004, priv. comm.).  Another example of a plate flaw on the POSS is K~2-4, discussed by Lutz \& Kaler (1983).

\section{Diagnostic Tools}\label{sec:diagnostic}

Since $>$90\% of PN have been discovered in the optical domain, we mainly discuss the principal optical diganostic plots in this review.   In general, diagnostic diagrams using a range of  emission-line intensities are useful for differentiating resolved emission sources such as PN from \HII\ regions, SNRs and other objects (e.g. Sabbadin, Minello \& Bianchini 1977; Cant\'o 1981; Fesen, Blair \& Kirshner 1985; Riera, Phillips \& Mampaso 1990; Phillips \& Guzman 1998; Kennicutt et al. 2000; Riesgo-Tirado \& L\'opez 2002; Riesgo \& L\'opez 2006; Kniazev, Pustilnik \& Zucker 2008; Raga et al. 2008).  However, the situation becomes more complex when compact or pseudo-stellar PN are considered.  There is also a rich literature concerning diagnostic plots and abundance determinations of extragalactic HII regions and active galactic nuclei (Veilleux \& Osterbrock 1987; Kewley \& Dopita 2002, Kewley et al. 2006, and references therein).  Discussion of the diagnostic parameters for these objects falls outside the scope of this paper.


The symbiotic stars for example have a considerable overlap with PN spectroscopically.  However, since symbiotic cores are generally denser than even the youngest PN, Guti\'errez-Moreno, Moreno \& Cort\'es (1995) have used the [O\,{\sc iii}] $\lambda$4363/$\lambda$5007 ratio to separate PN from medium- and high-excitation symbiotic stars.  This ratio is a good temperature indicator at densities typical of PN, and a good density indicator at the densities characteristic of symbiotics.  Therefore, diagnostic diagrams using these ratios can help differentiate true PN from most symbiotic stars.  Raman-scattered \OVI\ emission lines at $\lambda\lambda$6825, 7082\AA\ are a very useful diagnostic for symbiotic stars when visible (e.g. Schmid 1989).  S-type and D-type symbiotic stars can be further differentiated using the strengths of the HeI emission lines (Proga, Mikolajewska \& Kenyon 1994).  

However, we note that the overlap of properties between symbiotics and both pre-PN and compact PN means that taxonomic classification can be arbitrary.  For example, there is a continuum of observable properties from the symbiotic star CH~Cygni through~MWC 560 (V694~Mon) to the putative pre-PN, M~1-92 (Mikolajewski, Mikolajewska \& Tomov 1996; Arrieta, Torres-Peimbert \& Georgiev 2005).  This latter object may possibly be a symbiotic outflow.  Hb~12 is another interesting case, with properties intermediate between symbiotics and PN (Kwok \& Hsia 2007; Vaytet et al. 2009).

Schmeja \& Kimeswenger (2001) have used DENIS near-IR colour-colour plots to differentiate symbiotic stars (including those with resolved outflows) from compact PN.  Santander-Garcia, Corradi \& Mampaso (2009) have added pre-PN to this plot, showing that they largely overlap with PN and are separated from the symbiotic stars.  Corradi et al. (2008, 2009) have utilised various diagnostic plots combining IPHAS and 2MASS colours, which helps to effectively differentiate symbiotic stars from both normal stars and other \ha\ point-source emitters, including T~Tauri stars, Be stars, and CVs, as well as compact PN, when such objects have detections across all wavebands.  Mention must be again made of the IRAS two-colour diagram (e.g. van der Veen \& Habing 1988; Pottasch et al. 1988; G\'orny et al. 2001; Su\'arez et al. 2006) for classifying PN of various types, pre-PN, and post-AGB and AGB stars.  Similar plots utilizing MSX data have been devised by Ortiz et al. (2005).

Since optical emission-line ratios are the most widely utilised at the present time, a few caveats relevant to their use are given here:
\begin{itemize}
\item{When calculating line ratios, it is best to use lines of similar wavelengths, as reddening errors are then minimized, especially for PN candidates close to the Galactic plane or in the bulge.}

\item{In the blue, a high \OIII/\hb\ ratio is widely used as a PN diagnostic, but beware of the many mimics with F(5007) $\geq$ F(\hb) (see \S\ref{sec:mimics}).  Readers should also be aware of VLE PN with weak or undetectable \OIII\  lines.  The detection of \HeII\,$\lambda$4686 is a quite robust PN marker for resolved nebulae, with rare exceptions such as the \HeIII\ region G2.4+1.4 around the massive WO1 star WR~102; similar examples are also known in external galaxies (Pakull 2009) along with rare \HeIII\ regions ionized by luminous X-ray sources (e.g. Pakull \& Angebault 1986; see also Rappaport et al. 1994).  Note that the \HeII/\hb\ ratio is not a diagnostic for unresolved objects, as symbiotic stars commonly show $\lambda$4686 emission.}

\item{If red spectral data are available, both the \NII/\ha\ and \SII/\ha\ ratios should be investigated to better ascertain the nature of an emission nebula (e.g. Sabbadin, Minello \& Bianchini 1977; Cant\'o 1981, Riesgo \& L\'opez 2006).  In general, the \NII/\ha\ ratio in a photoionized nebula is sensitive to the nitrogen abundance, the hardness of the radiation field and is also influenced by the presence of shocks.  The dependence of the \SII/\ha\ ratio to shock conditions is well known, and has been frequently used to identify a likely SNR origin for the optical emission when \SII/\ha\ ratio exceeds 0.5 (e.g. Fesen et al. 1985), though several evolved PN have ratios that \emph{exceed} this value (e.g. Sh~2-188; Rosado \& Kwitter 1982).  An SNR identification is further supported by the presence of strong \OI, \OII, \NI\ and \NII\ lines.} 

\item{If no \SII\ emission is present, the [NII]/\ha\ ratio alone must be used with caution; values less than 0.6 could be either \HII\ regions or PN.  Note that very high excitation PN may have no measurable \NII\ or \SII\ emission, so other multi-wavelength data, as well as an investigation of the morphology, central star characteristics, and local environment, are needed to classify candidate high-excitation objects unambiguously.}
\end{itemize}

Based on a review of the literature, spectral-line data from our MASH catalogues, the SEC (Acker et al. 1992), as well as our large unpublished database of 450 emission-line sources rejected from MASH (Parker et al. 2010, in preparation), we offer a substantially revised and updated log F(\ha/\NII) versus log F(\ha/\SII) (SMB) diagram as a diagnostic tool (Figure~\ref{N2S2_plot}).  Line fluxes for Galactic \HII\ regions, SNRs, WR shells, and other objects are taken from an extensive unpublished database compiled from our own and literature data, to be analysed in full at a later date.  Line fluxes for LMC SNRs have been mainly taken from Payne, White \& Filipovi\'c (2008, and references therein).  For line-flux data on extragalactic \HII\ regions we primarily refer to Viironen et al. (2007, and references therein), supplemented with data for $\sim$310 metal-poor emission-line galaxies from Data Release 3 of the Sloan Digital Sky Survey (see Izotov et al. 2006). 

As a result of our extensive compendium of flux data for many different classes of emission-line sources, we note considerable overlap between domains, not previously appreciated, especially between \HII\  regions and low-excitation PN, between some evolved PN and SNRs, and between PN and LBV/WR ejecta nebulae.    In summary, we find that:

\begin{itemize}
\item PN contain various degrees of CNO-processed gas, so most are separated from ordinary \HII\ regions.  However, many Galactic \HII\ regions fall in the enlarged PN domain, and scatter over a much larger area than shown on previous versions of this plot.  The location of \HII\ regions in this plot are influenced by both metallicity and excitation parameter (e.g. Viironen et al. 2007).  The strong metallicity effect is easily seen by comparing the positions of the Galactic and extragalactic \HII\ regions; the latter sample is dominated by \HII\ regions in low-metallicity dwarf galaxies. 

\item Type~I PN (as defined by Kingsburgh \& Barlow 1994) are strongly enriched in nitrogen (and often helium).  Known Type~I PN were found to fall only in the bottom-left of the diagram, so we tentatively suggest a new domain, the realm of Type I objects.  PN with only red spectral data can be given a preliminary classification on the basis of this plot, prior to a full abundance analysis  (cf. Jacoby \& De Marco 2002).   These revised field boundaries are broadly consistent with PN photoionization modelling (Tajitsu et al. 1999), and with the data presented by Perinotto \& Corradi (1998). 

\item SNRs are systematically offset from the majority of  PN and \HII\ regions, though there is some overlap as signatures of shocks are sometimes found in PN and \HII\ regions, as well as some symbiotic and [Be] outflows.  Very old SNRs have optically emitting filaments that are dominated by swept-up ISM, and show abundances as predicted from modelling of shocks of the order of 100~\kms\ in ISM gas (Shull \& McKee 1979).  A few old interacting PN are travelling a similar speeds through the ISM, and have line ratios that can overlap with senile SNRs. However, on the whole, PN are more N-rich, reflecting CNO processing by the progenitor star, and plot systematically below the SNRs in the SMB diagram.

\item Magellanic Cloud (MC) SNRs are less metal-rich than Galactic SNRs, and systematically plot above their Galactic counterparts.

\item  Most shock-excited Herbig-Haro objects are clearly separated from SNRs, due to differing physical conditions, but there is considerable overlap between the fields.  Most of the bowshock nebulae discussed in \S~\ref{sec:miscell} (not plotted) overlap with the SNR domain.   

\item  WR shells and SNRs are largely separated in this plot.  The Galactic WR shells systematically plot below the MC WR shells, again due to metallicity --- most of the larger wind-blown shells plot amongst the ordinary HII regions, reflecting their composition which is dominated by swept-up local ISM.  Several Galactic WR nebulae are dominated by CNO-processed ejecta from their ionizing stars, and so plot in the PN domain, overlapping with the LBV-ejecta nebulae.  Some of the latter overlap with the Type I PN, reflecting very strong N-enrichment. 

\item Symbiotic stars and their resolved outflows generally plot in the PN field.  Note that many symbiotic stars cannot be plotted on the SMB diagram due to the absence of detectable \SII\ emission.  This is probably due to collisional quenching of the red \SII\ lines as a result of the characteristically high densities in these objects.

\end{itemize}

We reiterate that only Type~I PN can have the \emph{very strong} \NII\ emission (i.e. \NII/\ha\ $\geq$ 6) seen in some nebulae (Corradi et al. 1997; Kerber et al. 1998; Tajitsu et al. 1999; Frew, Parker \& Russeil 2006), with the exception of a few individual knots in very young SNRs like the Crab Nebula (MacAlpine et al. 1996) and a handful of LBV ejecta such as the outer knots around $\eta$~Carinae (e.g. Smith \& Morse 2004).   

We also present `BPT' (Baldwin, Phillips \& Terlevich 1981) diagrams using emission-line fluxes from our database.  Figure~\ref{BPT_plot} shows two versions of this plot: (a) log\,$F$($\lambda$5007)/$F$(\hb) versus log\,$F$($\lambda$6584)/$F$(\ha) and (b) log\,$F$($\lambda$5007)/$F$(\hb) versus log\,$F$(\SII)/$F$(\ha).  The overlap between fields is substantial, and while PN dominate at high $\lambda$5007/\hb\ ratios, many low-excitation PN populate the \HII\ regions of the diagrams.  WR shells can also have very high $\lambda$5007/\hb\ ratios.  While these diagrams are useful for PN researchers (e.g. Kniazev, Pustilnik \& Zucker 2008), they have more power in separating AGN from star-forming galaxies (Kewley et  al. 2006).

\begin{figure*}
\begin{center}
\includegraphics[scale=0.54,angle=-90]{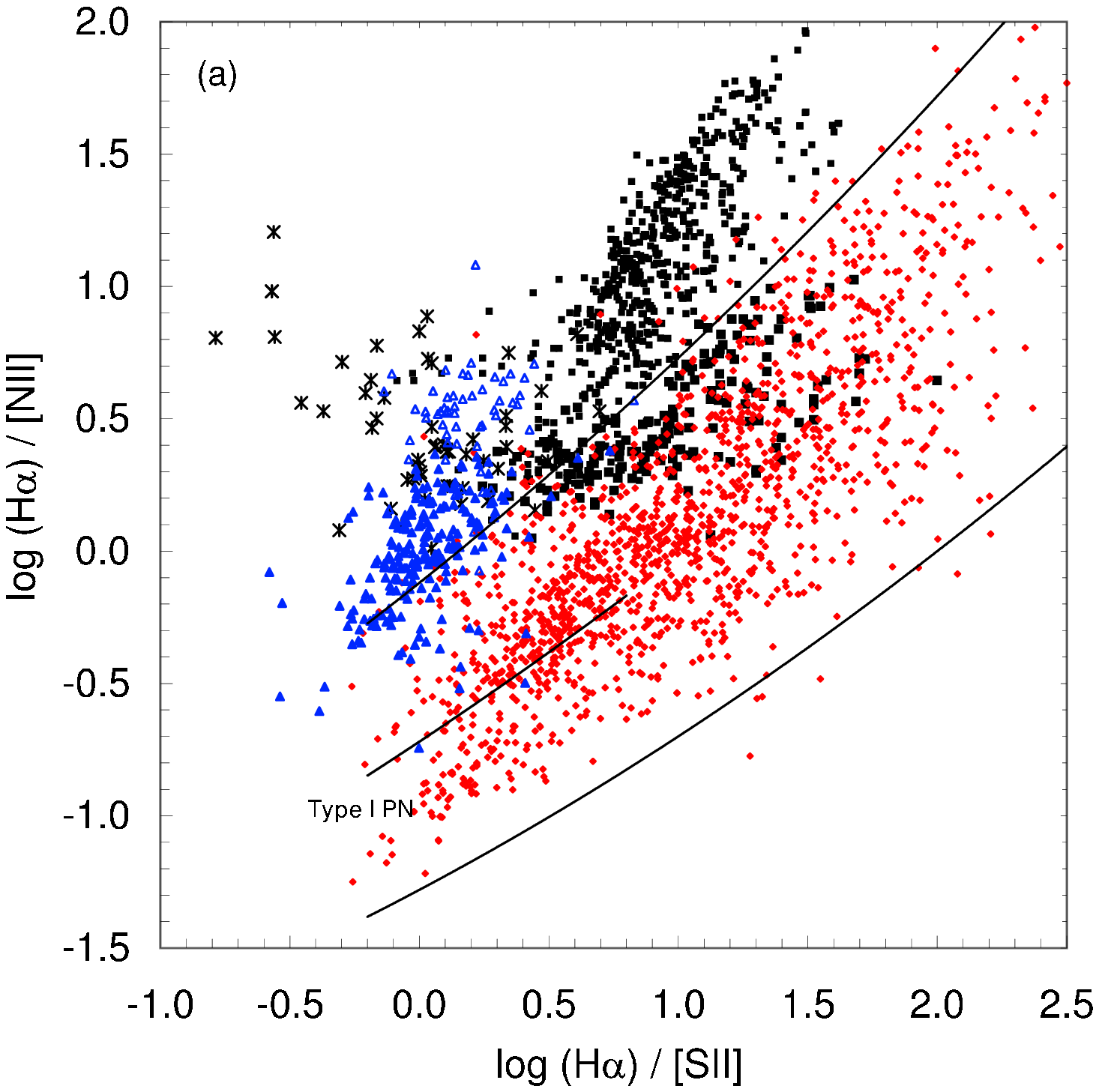}
\includegraphics[scale=0.54,angle=-90]{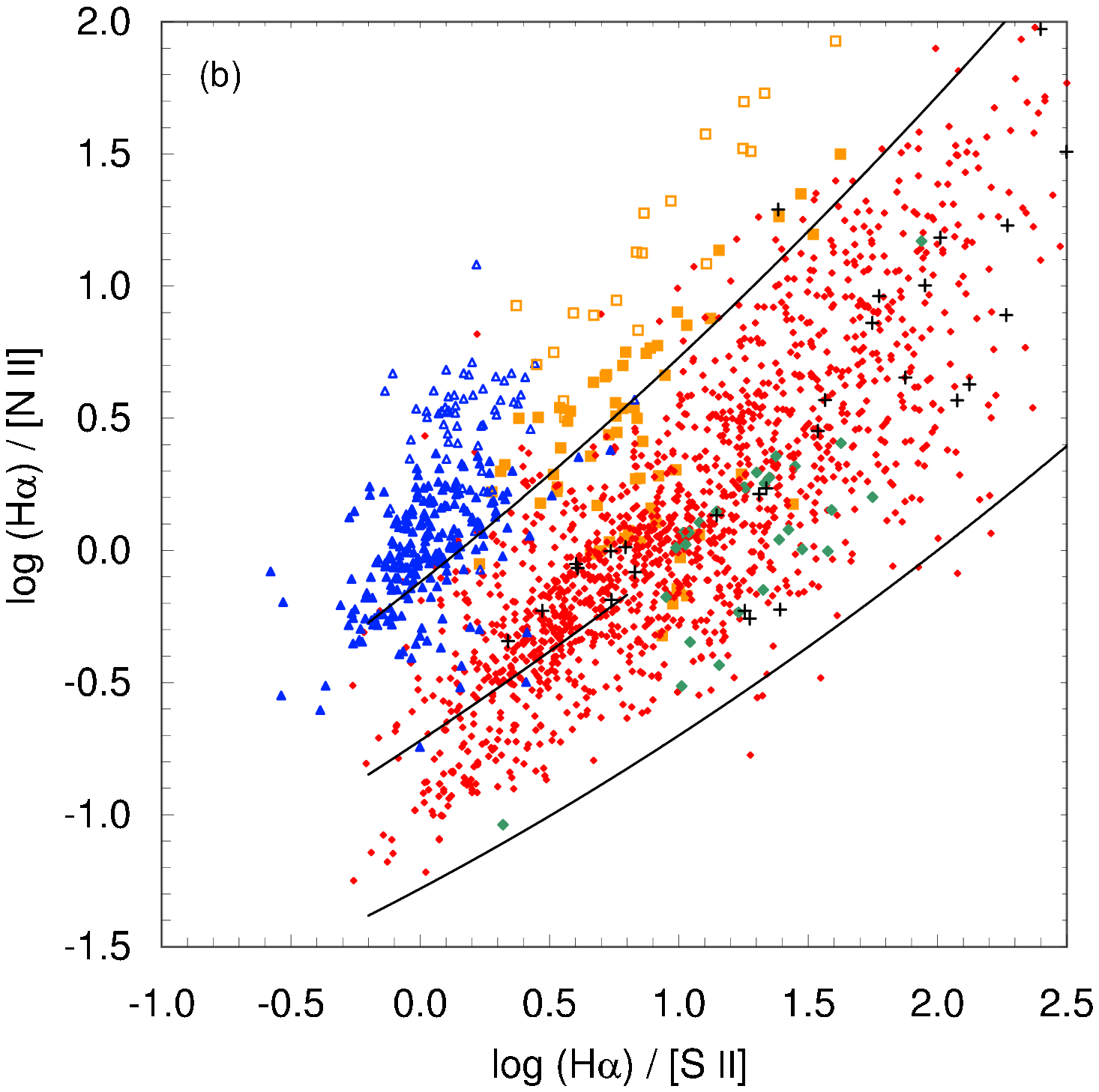}
\caption{(a) A revised version of the SMB or log\,$F$(H$\alpha$)/$F${[\rm N\sc ii]} versus log\,$F$(H$\alpha$)/$F${[\rm S\sc ii]} diagnostic diagram, where $F${[\rm N\sc ii]} refers to the sum of the two red nitrogen lines at $\lambda\lambda$6548, 6584\AA, and $F${[\rm S\sc ii]} refers to the sum of the two red sulfur lines at $\lambda\lambda$6717, 6731\AA.  The empirical boundaries for the PN field are modified here, based in part on new line fluxes from our unpublished database.  Galactic PN are plotted as red dots, HH regions as asterisks, Galactic SNRs as filled blue triangles, Magellanic Cloud SNRs as open blue triangles, Galactic \HII\ regions as large black squares, and extragalactic HII regions as small black squares.  We also plot a new domain, the realm of Type~I PN (as defined by Kingsburgh \& Barlow 1994). (b) The same plot as (a), showing in addition, symbiotic stars/outflows plotted as crosses, Galactic WR shells as orange squares, Cloud WR shells as open squares, and Galactic LBV nebulae as green diamonds.}
\label{N2S2_plot}
\end{center}
\end{figure*}

\begin{figure*}
\begin{center}
\includegraphics[scale=0.54,angle=-90]{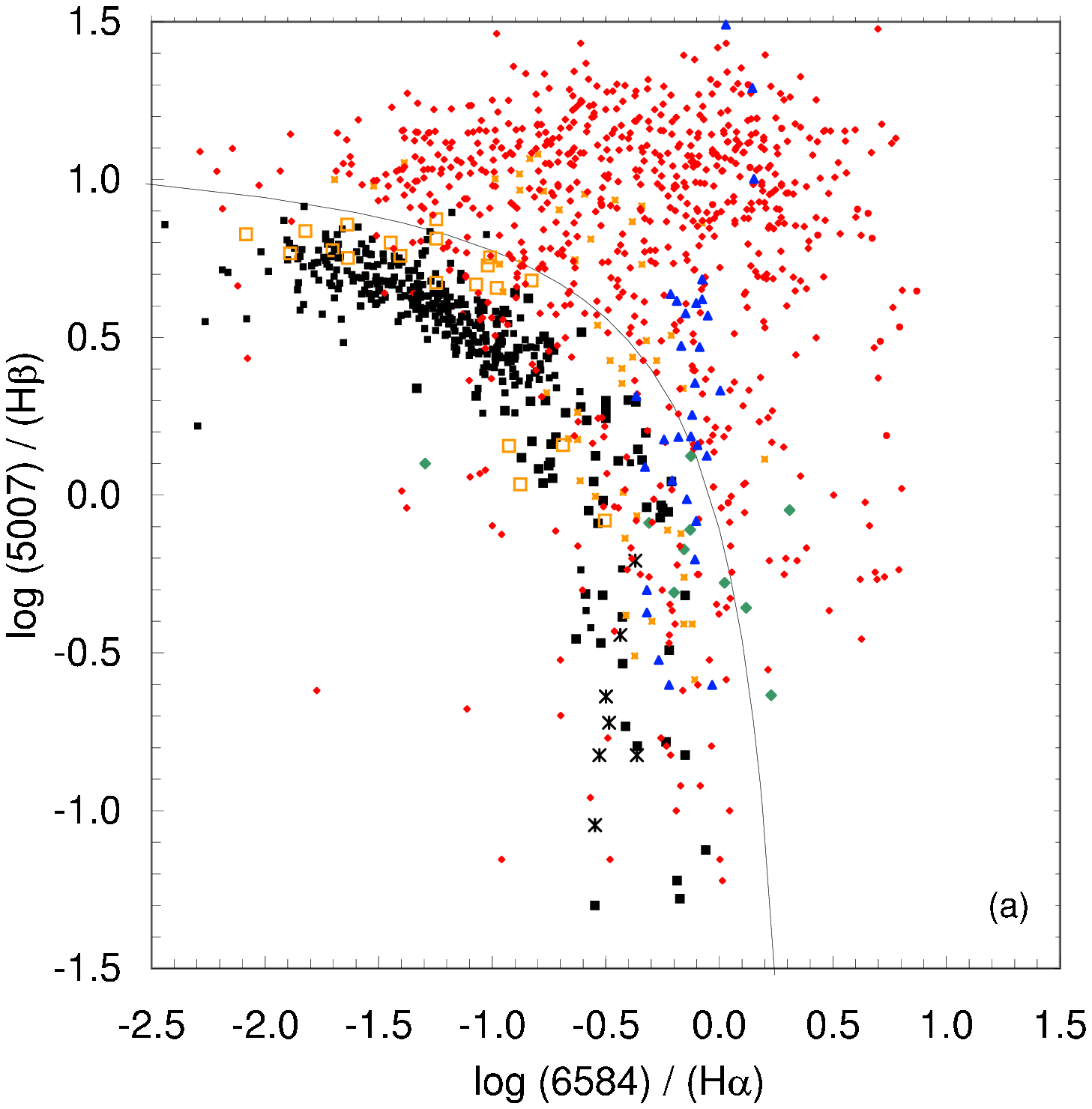}
\includegraphics[scale=0.54,angle=-90]{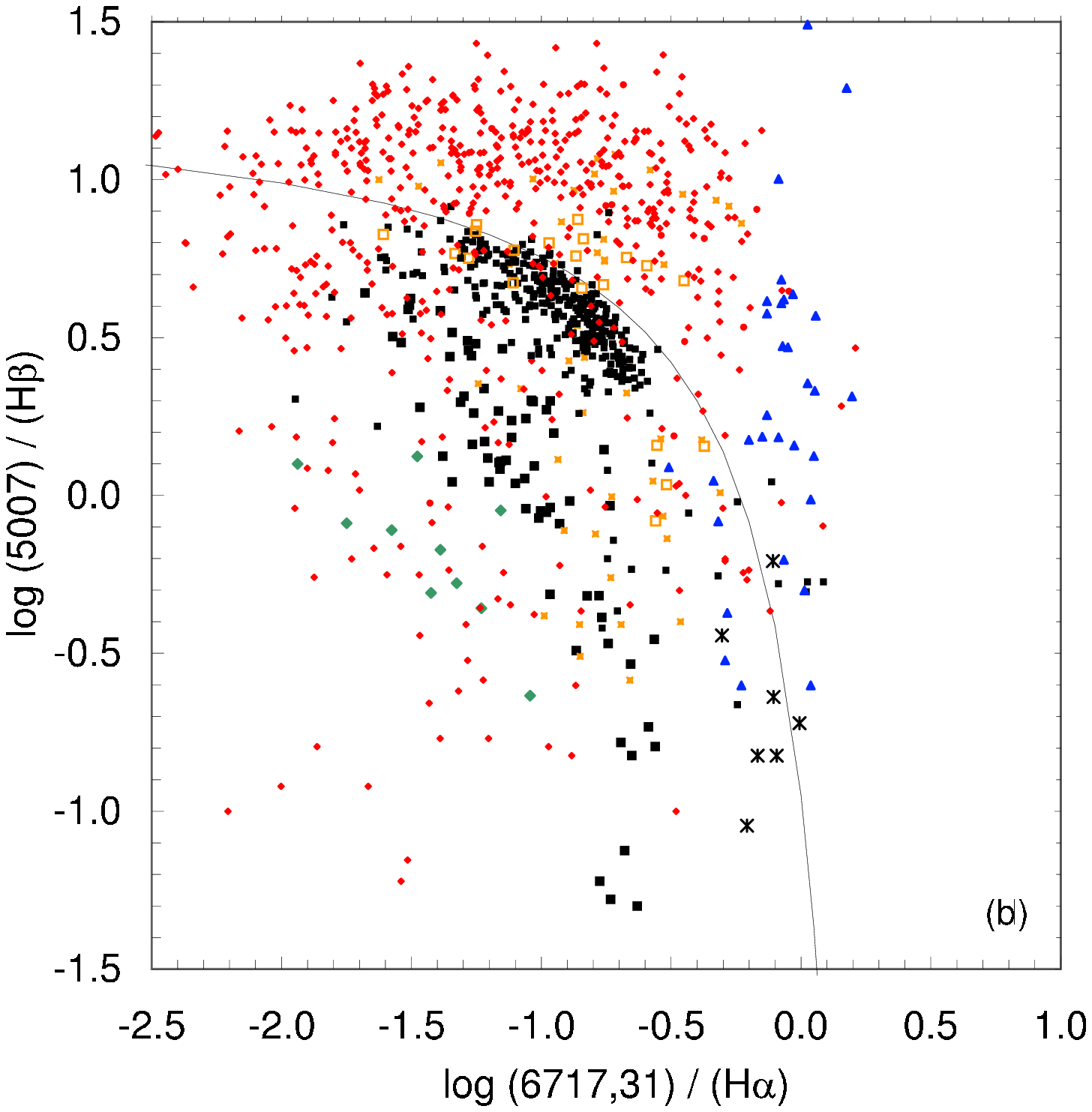}
\caption{(a) `BPT' log\,$F$(5007)/\hb\ versus log\,$F$(6584)/$F$(H$\alpha$) diagnostic diagram. (b) `BPT' log\,$F$(5007)/\hb\ versus log\,$F${[\rm S\sc ii]}/$F$(H$\alpha$) diagnostic diagram.  The curved boundaries are the classification curves from Kewley et al. (2006), separating star-forming galaxies (dominated by \HII\ regions) below and left of the line, from AGN.   Symbols as in Figure~\ref{N2S2_plot}.}
\label{BPT_plot}
\end{center}
\end{figure*}

\section{Weeding out the impostors}\label{sec:impostors}  

The ideal recipe to determine whether a candidate object is a true PN is based on a multi-wavelength approach applied to the morphological and spectroscopic characteristics of the nebula and its ionizing star.   This was previously difficult to achieve in most cases without detailed individual follow-up.  However, the recent advent of major wide-area surveys from the X-ray to the radio has improved the situation considerably.  Planned optical surveys such as VPHAS and Skymapper in the south will add to the already impressive data sets available under the framework of the Virtual Observatory.  It is the online availability and federation of these massive data sets that facilitates evaluation and discrimination of object types.  These surveys offer unprecedented opportunities for discovery (e.g. Parker et al. 2006; Miszalski et al. 2008; Viironen et al. 2009a, 2009b; Sabin et al. 2010), and fresh opportunities to eliminate contaminants that have plagued previous catalogues.   

We list here the various criteria that are used in our classification process, the exact combination depending on whether the candidate nebula is compact or extended.  Overall, we examine the:

\begin{itemize}
\item{Presence of a hot (blue) CS that is relatively faint compared to the nebular flux.  Non-blue CS in regions of low extinction may be binary stars; in such cases, a UV excess (e.g. found using GALEX data) can provide evidence for a hotter ionizing star.}
\item{Properties of this ionizing star, if known, including its evolutionary age and position in the HR diagram.}
\item{Nebular morphology: does the candidate PN have a typical shape, or is it irregular or otherwise pathological?}
\item{Nebular emission-line ratios, using diagnostic plots where applicable (see \S\ref{sec:diagnostic}).}
\item{Ionization structure, including the consistency of any ISM interaction with the proper-motion vector of ionizing star.}
\item{Near-IR and mid-IR colours, based on 2MASS, DENIS, IRAS, MSX and GLIMPSE data; use a range of diagnostic plots).}
\item{Available time-domain photometric data to search for variability of the ionizing star; this is useful for identifying symbiotic stars or close-binary CSPN.} 
\item{Strength of the radio and MIR flux densities (especially useful if a constraint on the distance is available).}
\item{Systemic velocity of nebula, and does it differ from the velocity of the ionizing star?}
\item{Line width of nebular gas: is the nebula expanding or is the line width (at FWHM) consistent with an \HII\ region or static ISM?  Note, however, that some PN can have low expanison velocities (2$v_{\rm exp}$ $<$ 20 \kms).  In compact or unresolved objects, extremely broad \ha\ wings are seen in young PN, symbiotic stars, and B[e] stars (Arrieta \& Torres-Peimbert 2003; Zickgraf 2003).}
\item{Abundances of the nebular gas --- check for products of core-He burning, which would indicate either a PN, a nova shell, or possibly massive star ejecta.} 
\item{Physical nebular diameter, which should be $\leq$ 5\,pc but need an estimate of the distance.  For example, bipolar symbiotic outflows are often much larger than bipolar \emph{pre-PN}.}
\item{Shklovsky (ionized) mass, which should be in the range 0.005 to 3\,$M_{\odot}$, but again need an estimate of the distance.}
\item{Local environment. For example, YSOs, T~Tauri stars and HAeBe stars are usually associated with \HII\ regions, molecular clouds, and areas of heavy obscuration or dark lanes.}
\item{Galactic latitude.  PN have a much larger scale height (250 $\pm$ 50\,pc; Zijlstra \& Pottasch 1991) than compact \HII\ regions, SNRs and massive stars, and are more likely to be found away from the Galactic plane than these objects.} 

\end{itemize}

No one criterion is generally enough to define the status of a candidate nebula, so we use the {\sl overall body of evidence} which is sufficiently compelling in most cases that we are confident in our interpretation.  As an illustrative example of our multiwavelength approach, the emission nebula around the sdB star PHL~932 is discussed in detail by Frew et al. (2010) in this issue.  However, a few objects refuse to be pigeonholed, and only further detailed investigations (and theoretical advances) will reveal their nature.



\section{Do PN form a heterogeneous class?}\label{sec:hetero}

In Section~\ref{sec:PN_definition}, we suggested a working phenomenological definition for a PN, and we reviewed the zoo of objects which can mimic PN in Section~\ref{sec:mimics}.  However, even after removal of the obvious mimics from PN catalogues, there is increasing evidence that the class itself is a `mixed bag' (e.g. G. Jacoby, 2006, quoted by De Marco 2009; Frew 2008) and that several stellar evolutionary pathways may produce nebulae best catalogued as PN.  A review of the literature indicates that PN-like nebulae (in other words, nebulae that meet the phenomenological criteria outlined in \S\ref{sec:PN_definition}) may arise from the following evolutionary channels:
\\

\noindent{\bf 1.  Post-AGB evolution of a single star} (or member of a wide, non-interacting binary), producing a conventional or classical PN.\\

\noindent{\bf 2(a). Long-period interacting binaries}, a subset of which appear to be related to the family of highly-collimated bipolar nebulae like M~2-9, He~3-401, and Mz~3, which host central stars exhibiting the B[e] phenomenon, and/or symbiotic characteristics (see \S\ref{sec:Be_stars}).\\

\noindent{\bf 2(b). Short-period interacting binaries}, namely those systems undergoing a common-envelope (CE) phase on the AGB, or possibly on the RGB, where close binarity is a prerequisite (Iben \& Livio 1993; De Marco 2009).  There is good evidence that these nebulae have quite distinct characteristics (Bond \& Livio 1990; Frew \& Parker 2009; Miszalski et al. 2009b).  We note that some post-CE PN appear to contain two subdwarf stars, though the exact evolution of these systems is currently unclear, as are the formation mechanism(s) of the putative PN around the classical novae GK~Per and V458~Vul (see \S\ref{sec:CVs}). \\

\noindent{\bf 3. The so-called `born-again' phenomenon}, where a final helium flash produces a H-deficient star and surrounding H-deficient knots inside a pre-existing, old PN (e.g. Sakurai's star).  Only a few examples are known (see Zijlstra 2002 for a review), but most of the outer shells have near-round morphologies (Kimeswenger et al.  2009).  It is possible that the H-deficient PN seen in the globular clusters M~22 and Fornax GC~5  (e.g. Gillett et al. 1989; Larsen 2008) represent evolved examples of the inner nebulae seen in Abell~30, Abell~58, Abell~78, and Sakurai's star, the outer nebulae having long since dissolved into the ISM.  

However, there may be problems with the standard interpretation of the born-again phenomenon.  Wesson (2008a) found a C/O ratio of less than unity and substantial quantities of neon in the ejecta of both Abell~30 and Abell~58, which are not predicted by very late thermal pulse models. The ejecta abundances are more akin to that seen in neon-rich novae! \\

\noindent{\bf 4. Scenarios that produce O(He) stars} (Rauch, Dreizler \& Wolff 1998).  Rauch et al. (2008) suggest two possible evolutionary scenarios for the formation of these rare stars.  They are possibly the long-sought successors of the R~CrB stars, which may result in turn from a double-degenerate merging process (Clayton et al. 2007).  Alternatively, the O(He) stars might be peculiar post early-AGB stars.  K~1-27 is the epitome of the class, but at any reasonable distance (Rauch, K\"oppen \& Werner 1994), this PN is very underluminous, falling off the main locus of PN in \ha\ surface brightness -- radius space (Frew \& Parker 2006; Frew 2008).  The discrepancy becomes worse if the closer distance of Ciardullo et al. (1999) is adopted.  

We also mention here the so-called hot R~CrB stars (De Marco et al. 2002), a very small collective of only four stars.  It appears that these too are a mixed bag, where the stars are without a common evolutionary history.  Two galactic stars, DY~Cen and MV~Sgr, have circumstellar nebulae but also have typical R~CrB helium abundances, which are not consistent with any post-AGB evolutionary models.  On the other hand, V348~Sgr, surrounded by a resolved PN-like nebula, and HV~2671 in the LMC have properties consistent with them being derived from a born-again evolutionary scenario.\\ 

\noindent{\bf 5. AGB-manqu\'e evolution}, potentially producing an ejecta nebula around an EHB star, e.g. PHL~932, but for a rebuttal, see Frew et al. (2010, in this issue).\\

\noindent{\bf 6. Evolution of a super-AGB star} (Herwig 2005; Poelarends et al. 2008), but there are no documented examples known.  However, Prieto et al. (2009) showed that an optical transient seen in the nearby spiral galaxy NGC~300 shares many properties with some bipolar pre-PN (as well as with SN 2008S in NGC~6946).  These authors conclude that an explosive event on a massive (6--10 M$_{\odot}$) carbon-rich AGB/super-AGB or post-AGB star is consistent with all available data, possibly producing a massive pre-PN,   but see Gogarten et al. (2009) and Smith et al. (2009) for contrary opinions.  In addition, the peculiar nebula LMC N66 surrounds a [WN4-5] star and has been suggested to be the product of a high-mass progenitor star, but a binary evolution channel may be more likely in this object (see Hamann et al. 2003 for a discussion).  The similar ([WN6]) Galactic object PM~5 (Morgan, Parker \& Cohen 2003) also needs detailed study.

Filipovi\'c et al. (2009) has recently drawn attention to some radio-luminous PN candidates in the Magellanic Clouds which appear brighter than expected.  These authors tentatively call the brighter objects `Super PNe', but we urge caution before such a moniker is used, especially in the absence of high-resolution optical imagery for some of their objects.  In fact, one of their PN candidates, SMC-N68, has been classified as a compact \HII\ region by Charmandaris, Heydari-Malayeri \& Chatzopoulis (2008; see also Meynadier \& Heydari-Malayeri 2007).  While PN in external galaxies have the advantage of being co-located at a common (and usually known) distance, the potential for contaminants in these samples is not trivial, and the problem warrants further investigation.  More work is needed to see if the bright end of the radio PN luminosity function is variable between different systems. The potential of symbiotic nebulae to contaminate the bright end of the PNLF is also largely unknown (see Frankowski \& Soker 2009, for a discussion).\\

\noindent{\bf 7. Exotica}.  This is a grab-bag of objects that defy straightforward classification.  We arbitrarily group two bipolar nebulae together first because of their extreme sizes (major axes of 5--7~pc) and unknown evolutionary origins; both have smaller putative PN within.  The first is He~2-111 (Figure~\ref{He2-111}), a strongly N-enhanced Type~I object (Perinotto \& Corradi 1998), surrounded by a huge, rapidly expanding `S-shaped' bipolar outflow (Webster 1978).  The strong nitrogen enrichment suggests a massive progenitor star.  The other object is KjPn~8 (L\'opez, V\'azquez \& Rodr\'iguez 1995) which also has a vast, point-symmetric, shock-excited bipolar outflow surrounding a compact PN-like core, which appears to have a Type~I chemistry; note that this core is very small and underluminous at the accepted distance of 1.6~kpc (Meaburn 1997), again falling way off the \ha\ SB-$r$ relation (Frew \& Parker 2010, in preparation).  L\'opez et al. (2000) speculate that the formation of KjPn 8 is due to two distinct and consecutive PN-like events, possibly resulting from a binary evolution channel.  This pathway may also be applicable to He~2-111, but until the CS is identified and studied, the status of this nebula remains unclear.

Another odd object is IPHAS J195935.55+283830.3 (Corradi et al. 2009), a peculiar He-rich emission-line star surrounded by a large 9\arcmin $\times$6\arcmin\ emission nebula.  These authors hypothesise that the compact source may be a photoionized, high density cocoon around a WR star.  We agree that it is most likely to be a Population I massive star, but a classification as a peculiar young PN cannot be ruled out at this stage.  

We also note a potential PN associated with the luminous OH/IR star V1018~Sco (Cohen, Parker \& Chapman 2005), a star that is unequivocally still in its AGB phase.  The origin of this ionized nebula is uncertain, but it may result from shock excitation.   

Lastly for completeness, we mention the old `nova' CK~Vul (Nova Vul 1670), only because it is surrounded by compact, expanding ejecta, as well as a very faint, larger bipolar emission nebula.  Several ideas have been proposed for its origin, none entirely satisfactory.  It has been suggested in the literature to be a slow nova, a young PN, a late thermal pulse object, or a stellar merger (see Hajduk et al. 2007, and references therein).  The merger of two main-sequence stars (Soker \& Tylenda 2003; Kato 2003), perhaps analogous to the eruptive variable V838~Mon, is probably the best hypothesis in our opinion.  A late-flashing AGB star seems ruled out on the basis of CK Vul's high luminosity in outburst ($M_{\rm V}$ $\sim$ $M_{\rm Bol}$ $\leq$ $-7$) at any realistic distance (Shara, Moffat \& Webbink 1985; Hajduk et al. 2007), so it seems unrelated to the PN fraternity.  However, the formation mechanism of the outer bipolar nebula remains unknown.

\begin{figure}[h]
\begin{center}
\includegraphics[scale=0.2]{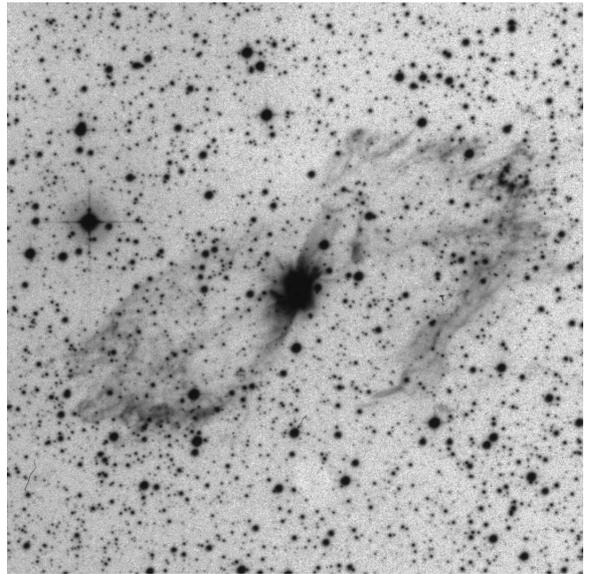}
\caption{SHS (Parker et al. 2005) H$\alpha$+\NII\ image of He~2-111, a Type~I PN surrounded by a huge, point-symmetric bipolar outflow first noted by Webster (1978).  Image is 720\arcsec\ on a side with NE at top left.}
\label{He2-111}
\end{center}
\end{figure}

\subsection{Relative Numbers}
In summary, it is difficult to ascertain the relative contributions of these various evolutionary pathways to the PN population, but based on a volume-limited, solar neighbourhood sample, Frew \& Parker (2009) found 22 $\pm$ 9\,\% of CSPN are close binaries and are presumably derived from a post-CE evolutionary channel (De Marco 2009).  While preliminary, this number is in agreement with a binary-fraction estimate of 12--21\% for a sample of Galactic bulge PN (Miszalski 2009a).  

Within a census of 200+ PN within 2.0\,kpc of the Sun (Frew 2008), there are only one or two born-again objects (Abell~30 and Abell~78), one PN with an O(He) nucleus (K~1-27), one putative PN with a [WN] CS (PM~5), and a couple of collimated bipolar outflows associated with [Be] stars, e.g. M~2-9.  As for the fraction of PN holding unresolved high-density cores, both EGB~6 and NGC 6804 are within 1\,kpc, so the fraction of EGB~6-like CS is 2/55 or $\sim$4\%.  This is a very preliminary estimate, not only because of small number statistics, but because not all CS in the solar neighbourhood have adequate spectroscopy.  

At this stage we make no attempt to estimate the number of bona fide pre-PN in the local volume due to the poor distance information for many of these objects (an excellent catalogue of post-AGB objects and pre-PN is given by Szczerba et al. 2007).  We will present a more detailed statistical analysis utilising a larger volume-limited sample of 400+ PN in a forthcoming paper.

\section{Conclusions}
The total number of true, likely and possible PN now known in the Milky Way is nearly 3000, almost double the number known a decade ago, and with the potential to grow more over the next few years.  These discoveries are a legacy of the recent availability of wide field, narrowband, imaging surveys.  

Many types of objects can mimic PN, and consequently our current  Galactic and extragalactic samples are liberally salted with emission-line objects and other sources masquerading as PN.  
Many contaminants still remain to be excised as PN, particularly from the pre-MASH catalogues.   

On the plus side, we find that much improved discrimination of true PN from their mimics is now possible based on the wide variety of high-quality UV, optical, IR, and radio data that is now available.  We recommend several diagnostic plots for classification, but it pays to be aware of their limitations, as it does to be cognisant of the full zoo of potential mimics.  While data-mining has its place, we caution that unusual objects be classified on a case-by-case basis.

On the downside is the realisation that the overall `PN phenomenon' probably includes a melange of subgroups that observationally have subtle differences, though we think these are quite rare.  It appears likely that these subgroups have resulted from a range of discrete evolutionary channels.   In other words, the PN phenomenon appears to be a mixed bag.  In particular, the role of binarity in the formation and shaping of PN is still an open question (Soker \& Livio 1989; Soker 1997, 1998; Moe \& De Marco 2006; De Marco 2009; Miszalski et al. 2009b), but it appears likely that many CS appear to be single stars (Soker 2002; Soker \& Subag 2005).  We reiterate that there is a surprising diversity in the spectra of emission-line CS of PN-like nebulae, but the exact relationships between these nuclei and \emph{some} symbiotic and B[e] stars is still unclear.  Future observational and theoretical advances will allow a more precise taxonomy of the PN phenomenon, but a consensus eludes the community at this point in time.

\section*{Acknowledgements}

We thank the anonymous referee for constructive comments and Martin Cohen, Orsola De Marco, Simon O'Toole, Geoff Clayton, and Brent Miszalski for useful discussions.  We also thank Joe Stupar for kindly providing some SNR line-ratio data in advance of publication.   DJF gratefully acknowledges Macquarie University for an Australian Postgraduate Award, and also the Government of Western Australia for additional financial support.  This research has made use of the SIMBAD database, operated at CDS, Strasbourg, and has used data from the AAO/UKST H$\alpha$ Survey, produced with the support of the Anglo-Australian Telescope Board and the Particle Physics and Astronomy Research Council.  Additional data was used from the Southern H$\alpha$ Sky Survey Atlas (SHASSA) and the Wisconsin H-Alpha Mapper (WHAM), which were produced with support from the National Science Foundation.

\section*{References}

Abell, G.O. 1966, ApJ, 144, 259	\\	\\
Acker, A., Chopinet, M., Pottasch, S.R., \& Stenholm, B. 1987,  A\&AS, 71, 163	\\	\\
Acker, A., Stenholm B., V\'eron, P. 1991, A\&AS, 87, 499	\\	\\
Acker, A., Ochsenbein, F., Stenholm, B., Tylenda, R., Marcout, J., Schohn, C. 1992,  Strasbourg-ESO Catalogue of Galactic Planetary Nebulae (Garching: ESO)	\\	\\
Acker, A., Marcout J., Ochsenbein F. 1996,  First Supplement to the SECPGN (Observatoire de Strasbourg)	\\	\\
Allen, D.A. 1982, in The Nature of Symbiotic Stars, ed. M. Freidjung, R. Viotti (Dordrecht: D. Reidel), p. 27	\\	\\
Allen, D.A. 1984, PASA, 5, 369	\\	\\
Arrieta, A., Torres-Peimbert, S. 2003, ApJS, 147, 97	\\	\\
Arrieta, A., Torres-Peimbert, S., Georgiev, L. 2005, ApJ, 623, 252	\\	\\
Baldwin, J.A., Phillips, M.M., Terlevich R. 1981, PASP, 93, 5	\\	\\
Balick, B.,  Frank, A. 2002, ARA\&A, 40, 439	\\	\\
Beaulieu, S.F., Dopita, M.A.,  Freeman, K.C. 1999,  ApJ, 515, 610	\\	\\
Belczy\'nski K., Mikolajewska, J., Munari, U., Ivison, R.J., Friedjung, M. 2000,   A\&AS, 146, 407	\\	\\
Bica, E., Claria, J.J., Bonatto, C., Piatti, A.E., Ortolani, S., Barbuy B. 1995,  A\&A, 303, 747	\\	\\
Bilikova, J., Chu, Y.-H., Su, K., Gruendl, R., Rauch, T., De Marco, O., Volk, K. 2009, Journal of Physics: Conf. Ser., 172, (1), 012055	\\	\\
Blair, W.P., Kirschner, R.P.,  Fesen, R.A., Gull T.R. 1984, ApJ, 282, 161	\\	\\
Bode, M.F. 2004, in ASP Conf. Series, 313, Asymmetric Planetary Nebulae III, ed. M. Meixner, J. Kastner, N. Soker (San Francisco: ASP), p. 504	\\	\\
Bode, M.F. 2009, AN (in press), eprint: arXiv:0911.5254	\\	\\
Bode, M.F., Roberts, J.A., Whittet, D.C.B., Seaquist, E.R., Frail, D.A. 1987,  Nature, 329, 519	\\	\\
Bohigas, J., Tapia, M. 2003, AJ, 126, 1861	\\	\\
Bond, H.E. 2009, Journal of Physics: Conf. Ser., 172, (1), 012029	\\	\\
Bond, H.E., Livio, M. 1990, ApJ, 355, 568	\\	\\
Boumis, P., Paleologou, E.V., Mavromatakis, F., Papamastorakis, J. 2003,   MNRAS, 339, 735	\\	\\
Boumis, P., Akras, S., Xilouris, E.M., Mavromatakis, F., Kapakos, E., Papamastorakis, J., Goudis C.D. 2006, MNRAS, 367, 1551	\\	\\
Cant\'o, J., 1981, in Astrophysics and Space Science Library, volume 91, Investigating the Universe, ed. F.D. Kahn (Dordrecht: D. Reidel), p. 95	\\	\\
Cappellaro, E., Sabbadin, F., Salvadori, L., Turatto, M.,  Zanin, C. 1994,  MNRAS, 267, 871	\\	\\
Carey S.J. et al. 2009, PASP, 121, 76	\\	\\
Charmandaris, V., Heydari-Malayeri, M., Chatzopoulis, E. 2008, A\&A, 487, 567	\\	\\
Chu, Y.-H., Gruendl, R.A., Williams, R.M., Gull, T.R., Werner, K. 2004,  AJ, 128, 2357	\\	\\
Ciardullo, R. 2010, PASA, in press (this issue), eprint: arXiv:0909.4356	\\	\\
Ciardullo, R., Jacoby, G.H., Harris, W.E. 1991, ApJ, 383, 487	\\	\\
Ciardullo, R., Bond, H.E., Sipior, M.S., Fullton, L.K., Zhang, C.-Y., Schaefer, K.G. 1999, AJ, 188, 488	\\	\\
Clayton, G.C., Geballe, T.R., Herwig, F., Fryer, C., Asplund, M. 2007, ApJ, 662, 1220	\\	\\
Cohen, J.G., Rosenthal, A.J. 1983,  ApJ, 268,  689	\\	\\
Cohen, M., Parker, Q.A. 2003, in IAU Symp., 209, Planetary Nebulae: Their Evolution and Role in the Universe, ed. S. Kwok, M. Dopita, R. Sutherland (San Francisco: ASP), p. 33	\\	\\
Cohen, M., Parker, Q.A., Chapman, J. 2005, MNRAS, 357, 1189	\\	\\
Cohen, M., FitzGerald, M.P., Kunkel, W., Lasker, B.M., Osmer, P.S. 1978,  ApJ, 221, 151	\\	\\
Cohen, M., Parker Q.A., Green A. 2005, MNRAS, 360, 1439	\\	\\
Cohen, M.C. et al. 2005, ApJ, 627, 446	\\	\\
Cohen, M.C. et al. 2007, ApJ, 669, 343	\\	\\
Cohen, M.C. et al. 2010, in preparation	\\	\\
Copetti, M.V.F., Oliveira, V.A., Riffel, R., Casta\~neda, H.O., Sanmartim, D. 2007, A\&A, 472, 847	\\	\\
Corradi, R.L.M. 1995.  MNRAS, 276, 521  	\\	\\
Corradi, R.L.M. 2003, in ASP Conf. Proceedings, 303, Symbiotic Stars Probing Stellar Evolution, ed. R.L.M Corradi, J. Mikolajewska, T.J. Mahoney (San Francisco: ASP), p. 393	\\	\\
Corradi, R.L.M., Schwarz, H.E. 1993.  A\&A, 268, 714  	\\	\\
Corradi, R.L.M., Schwarz, H.E. 1995.  A\&A, 293, 871  	\\	\\
Corradi, R.L.M., Schwarz, H.E. 1997, in Physical Processes in Symbiotic Binaries and Related Systems, ed. J. Mikolajewska (Warsaw: Copernicus Foundation for Polish Astronomy), p. 147	\\	\\
Corradi, R.L.M., Villaver, E., Mampaso, A., Perinotto, M. 1997, A\&A, 324, 276   	\\	\\
Corradi, R.L.M., Brandi, E., Ferrer, O.E., Schwarz H.E. 1999, A\&A, 343, 841   	\\	\\
Corradi, R.L.M., Livio, M., Balick, B., Munari, U., Schwarz, H.E. 2001, ApJ, 553, 211  	\\	\\
Corradi, R.L.M.; Sch\"onberner, D., Steffen, M., Perinotto, M. 2003, MNRAS, 340, 417	\\	\\
Corradi, R.L.M. et al. 2008,  A\&A, 480, 409	\\	\\
Corradi, R.L.M. et al. 2009, A\&A, in press, eprint: arXiv:0910.5930	\\	\\
De Marco, O. 2008, in ASP Conf.  Ser., 391, Hydrogen-Deficient Stars, ed. K. Werner, T. Rauch (San Francisco: ASP), p. 209	\\	\\
De Marco, O. 2009, PASP, 121, 316	\\	\\
De Marco, O., Clayton, G.C., Herwig, F., Pollacco, D.L., Clark, J.S., Kilkenny, D. 2002, AJ, 123, 3387	\\	\\
Dengel, J., Hartl, H., Weinberger R. 1980,  A\&A, 85, 356	\\	\\
Dopita, M.A. 1978, ApJS, 37, 117	\\	\\
Downes, R.A., Duerbeck, H.W. 2000, AJ, 120, 2007	\\	\\
Drew, J.E. et al. 2005,  MNRAS, 362, 753	\\	\\
Egan, M.P., Clark, J.S., Mizuno, D., Carey, S., Steele, I.A., Price, S.D. 2002, ApJ, 572, 288 	\\	\\
Ellis, G.L., Grayson, E.T., Bond, H.E. 1984, PASP, 96, 283	\\	\\
Fesen, R.A., Blair, W.P., Kirshner, R.P., 1985, ApJ, 292, 29	\\	\\
Filipovi\'c, M.D. et al. 2009, MNRAS, 399, 769	\\	\\
Flagey, N. et al. 2009, BAAS, 41, 762	\\	\\
Frankowski, A., Jorissen, A. 2007,  Baltic Astron., 16, 104	\\	\\
Frankowski, A., Soker, N. 2009, ApJ, 703, L95	\\	\\
Frew, D.J. 2004, JAD, 10, 6	\\	\\
Frew, D.J. 2008, Unpublished PhD Thesis, Macquarie University	\\	\\
Frew, D.J., Parker, Q.A. 2006, in IAU Symp., 234, Planetary Nebulae in our Galaxy and Beyond, ed. M.J. Barlow, R.H. M\'endez (Cambridge: CUP), p. 49	\\	\\
Frew, D.J., Parker, Q.A. 2009, in Asymmetrical Planetary Nebulae IV, IAC Electronic Pub., ed. R.L.M. Corradi, A. Manchado, N. Soker (La Palma: Instituto de Astrof\'isica de Canarias), p. 475 	\\	\\
Frew, DJ., Madsen, G.J., Parker, Q.A. 2006, in IAU Symp., 234, Planetary Nebulae in our Galaxy and Beyond, ed. M.J. Barlow, R.H. M\'endez (Cambridge: CUP), p. 395	\\	\\
Frew, D.J., Parker, Q.A., Russeil, D. 2006, MNRAS, 372, 1081	\\	\\
Frew, D.J., Madsen, G.J., O'Toole, S.J., Parker, Q.A. 2010, PASA, in press; eprint: arXiv:0910.2078	\\	\\
Garc\'ia-Lario, P., Manchado, A., Pych, W., Pottasch, S.R. 1997,  A\&AS, 126, 479	\\	\\
Garc\'ia-Lario, P., Riera, A., Manchado, A. 1999, ApJ, 526, 854	\\	\\
Gaustad, J.E., McCullough, P.R., Rosing, W., Van Buren, D.J.  2001, PASP, 113, 1326	\\	\\
Gesicki, K., Zijlstra, A.A., Szyszka, C., Hajduk, M., Lagadec, E., Guzman Ramirez, L. 2010, A\&A, in press, eprint: arXiv:1001.5387	\\	\\
Gillett, F.C. et al. 1989, ApJ, 338, 862	\\	\\
Gogarten, S.M., Dalcanton, J.J., Murphy, J.W., Williams, B.F., Gilbert, K., Dolphin, A. 2009, ApJ, 703, 300	\\	\\
Gon\c{c}alves, D.R., Corradi, R.L.M., Mampaso, A. 2001, ApJ, 547, 302	\\	\\
Goodrich, R.W. 1991,  ApJ, 376, 654	\\	\\
G\'orny, S.K., Stasi\'nska, G., Szczerba, R., Tylenda, R. 2001, A\&A, 377, 1007	\\	\\
Greiner, J. et al. 2001,  A\&A, 376, 1031	\\	\\
Guerrero, M.A., Miranda L.F., Chu Y.-H., Rodr\'iguez, M., Williams, R.A. 2001, ApJ, 563, 883	\\	\\
Gussie, G.T. 1995, PASA, 12, 31	\\	\\
Guti\'errez-Moreno, A.,  Moreno, H. 1998,  PASP, 110, 458	\\	\\
Guti\'errez-Moreno, A., Moreno, H., Cort\'es, G. 1995,  PASP, 107, 462	\\	\\
Hajduk, M. et al. 2007, MNRAS, 378, 1298	\\	\\
Hajduk, M., Zijlstra, A.A., Gesicki, K. 2008, A\&A, 490, L7	\\	\\
Hamann, W.-R., Pe\~na, M., Gr\"afener, G., Ruiz, M.T.  2003, A\&A, 409, 969	\\	\\
Henize, K.G. 1967, ApJS, 14, 125	\\	\\
Herwig, F. 2005, ARA\&A, 43, 435	\\	\\
Hester, J.J., Kulkarni, S.R. 1988, ApJ, 331, L121	\\	\\
Hodge, P.W., Zucker, D.B. \& Grebel, E.K. 2000, BAAS, 197, 3812	\\	\\
Hoessel, J.G., Saha, A., Danielson, G.E. 1988, PASP, 100, 680	\\	\\
Hollis, J.M., Van Buren, D., Vogel, S.N., Feibelman, W.A., Jacoby, G.H., Pedelty, J.A. 1996, ApJ, 456, 644	\\	\\
Hrivnak, B.J., Smith, N., Su, K.Y.L., Sahai, R. 2008, ApJ, 688, 327	\\	\\
Iben, I., Jr.,  Livio, M. 1993, PASP, 105, 1373	\\	\\
Izotov, Y.I., Stasi\'nska, G., Meynet, G., Guseva, N.G., Thuan, T.X. 2006, A\&A, 448, 955	\\	\\
Jacoby, G.H. 1981, ApJ, 244, 903	\\	\\
Jacoby, G.H. 2006, in Planetary Nebulae beyond the Milky Way, ed. L. Stanghellini, J.R. Walsh, N.G. Douglas (Berlin: Springer), p.17	\\	\\
Jacoby, G.H., De Marco, O. 2002, AJ, 123, 269	\\	\\
Jacoby, G.H., Van de Steene, G. 2004, A\&A, 419, 563	\\	\\
Jacoby, G.H., Ciardullo, R., Ford, H.C. 1990, ApJ, 356, 332	\\	\\
Jacoby, G.H. et al. 2009, PASA, these proceedings (in press), eprint: arXiv:0910.0465	\\	\\
Jorissen, A. 2003, in ASP Conf. Ser. 303, Symbiotic Stars Probing Stellar Evolution, ed. R.L.M. Corradi, J. Mikolajewska, T.J. Mahoney (San Francisco: ASP), p. 25	\\	\\
Jorissen, A., Za\u{c}s, L., Udry, S., Lindgren, H., Musaev, F.A. 2005, A\&A, 441, 1135	\\	\\
Kahabka, P., van den Heuvel, E.P.J. 1997, ARA\&A, 35, 69	\\	\\
Kaler, J.B. 1981, ApJ, 245, 568	\\	\\
Kaler, J.B., Feibelman, W.A. 1985, PASP, 97, 660	\\	\\
Kastner, J.H., Weintraub, D.A., Zuckerman, B., Becklin, E.E., McLean, I., Gatley, I. 1992, ApJ, 398, 552	\\	\\
Kato, T. 2003, A\&A, 399, 695	\\	\\
Kennicutt, R.C., Jr, Bresolin, F., French, H. \& Martin, P. 2000, ApJ, 537, 589	\\	\\
Kenyon, S.J. 1986, The Symbiotic Stars (Cambridge: CUP)	\\	\\
Kerber, F., Roth, M., Manchado, A., Gr\"obner, H. 1998,  A\&AS, 130, 501	\\	\\
Kewley, L.J.,  Dopita, M.A. 2002, ApJS, 142, 35	\\	\\
Kewley, L.J., Groves B., Kauffmann G.,  Heckman T. 2006,  MNRAS, 372, 961	\\	\\
Kimeswenger, S. 1998,  MNRAS, 294, 312	\\	\\
Kimeswenger, S. et al. 2009, in Asymmetrical Planetary Nebulae IV, IAC Electronic Pub., ed. R.L.M. Corradi, A. Manchado, N. Soker (La Palma: Instituto de Astrof\'isica de Canarias), p. 125	\\	\\
Kingsburgh, R.L, Barlow, M.J. 1994, MNRAS, 271, 257	\\	\\
Kingsburgh R.L., English J. 1992,  MNRAS, 259, 635	\\	\\
Kistiakowsky, V., Helfand, D.J. 1995, AJ, 110, 2225	\\	\\
Kniazev, A.Y., Pustilnik, S.A., Zucker, D.B. 2008, MNRAS, 384, 1045	\\	\\
Kohoutek, L. 1963, BAC, 14, 70	\\	\\
Kohoutek, L. 2001, A\&A, 378, 843	\\	\\
Kraus, M., Borges Fernandes, M., De Ara\'ujo, F.X., Lamers, H.J.G.L.M. 2005, A\&A, 441, 289	\\	\\
Kraus, M., Borges Fernandes, M., Chesneau, O. 2009, eprint: arXiv:0909:2268	\\	\\
Kronberger, M. et al. 2006, A\&A, 447, 921	\\	\\
Kwok, S. 2000, The Origin and Evolution of Planetary Nebulae (Cambridge \& New York: Cambridge University Press)	\\	\\
Kwok, S. 2003, in ASP Conf. Proceedings, 303, Symbiotic Stars Probing Stellar Evolution, ed. R.L.M Corradi, J. Mikolajewska, T.J. Mahoney (San Francisco: ASP), p. 428	\\	\\
Kwok, S. 2010, PASA, in press (this issue), eprint: arXiv:0911.5571	\\	\\
Kwok, S., Hsia, C.H. 2007, ApJ, 660, 341	\\	\\
Kwok, S., Purton, C.R., Fitzgerald, P.M. 1978, ApJ, 219, L125	\\	\\
Kwok, S., Zhang, Y., Koning, N., Huang, H.-H., Churchwell, E. 2008, ApJS, 174, 426	\\	\\
Lachaume, R., Preibisch, Th., Driebe, Th., Weigelt, G. 2007, A\&A, 469, 587	\\	\\
Meynadier, F., Heydari-Malayeri, M. 2007, A\&A, 461, 565	\\	\\
Larsen, S.S. 2008, A\&A, 477, L17	\\	\\
Lauberts, A. 1982, The ESO/Uppsala Survey of the ESO (B) Atlas (Garching: European Southern Observatory)	\\	\\
Liebert, J., Green, R., Bond, H.E., Holberg, J.B., Wesemael, F., Fleming, T.A., Kidder, K. 1989, ApJ, 346, 251	\\	\\
Livio, M., Soker, N. 2001, ApJ, 552, 685	\\	\\
Longmore, A.J. 1977,  MNRAS, 178, 251	\\	\\
L\'opez, J.A., V\'azquez, R., Rodr\'iguez, L.F. 1995, ApJ, 455, L63	\\	\\
L\'opez, J.A., Meaburn, J., Rodr\'iguez, L.F., V\'azquez, R., Steffen, W., Bryce, M. 2000, ApJ, 538, 233	\\	\\
Lutz, J.H. 1984, ApJ, 279, 714 	\\	\\
Lutz, J.H., Kaler, J.B. 1983, PASP, 95, 739	\\	\\
Lutz, J.H., Kaler, J.B., Shaw, R.A., Schwarz, H.E., Aspin, C. 1989, PASP, 101, 966	\\	\\
MacAlpine, G.M., Lawrence, S.S., Sears, R.L., Sosin, M.S., Henry, R.B.C. 1996, ApJ, 463, 650	\\	\\
Madsen, G.J., Frew, D.J., Parker, Q.A., Reynolds, R.J., Haffner, L.M.  2006, in IAU Symp., 234, Planetary Nebulae in our Galaxy and Beyond, ed. M.J. Barlow, R.H. M\'endez (Cambridge: CUP), p. 455	\\	\\
Makarov, D.I., Karachentsev, I.D., Burenkov, A.N. 2003, A\&A, 405, 951	\\	\\
Mampaso, A. et al. 2006, A\&A, 458, 203	\\	\\
Marcolino, W.L.F., de Ara\'ujo, F.X. 2003, AJ, 126, 887	\\	\\
Marston, A.P. 1997, ApJ, 475, 188	\\	\\
Marston, A.P., McCollum, B. 2008, A\&A, 477, 193	\\	\\
Marston, A.P., Yocum, D.R., Garcia-Segura, G., Chu, Y.-H. 1994, ApJS, 95, 151	\\	\\
Martin-Hern\'andez, N.L., Esteban, C., Mesa-Delgado, A., Bik, A., Puga, E. 2008, A\&A, 482, 215	\\	\\
Meaburn, J. 1997, MNRAS, 292, L11	\\	\\
Meynadier, F., Heydari-Malayeri, M. 2007, A\&A, 461, 565	\\	\\
Mikolajewski, M., Mikolajewska, J., Tomov, T. 1996, in IAU Symp., 165, Compact Stars in Binaries, ed. J. van Paradijs, E.P.J. van den Heuvel, E. Kuulkers (Dordrecht: Kluwer) p. 451	\\	\\
Minkowski R. 1946,  PASP, 58, 305	\\	\\
Miroshnichenko, A.S. 2006, in ASP Conf. Series, 355, Stars with the B[e] Phenomenon, ed. M. Kraus, A.S. Miroshnichenko (San Francisco: ASP), p.13	\\	\\
Miroshnichenko, A.S. 2007, ApJ, 667, 497	\\	\\
Miszalski, B., Parker, Q.A., Acker, A., Birkby, J., Frew, D.J., Kovacevic, A. 2008, MNRAS, 384, 525	\\	\\
Miszalski, B., Acker, A., Moffat, A.F.J., Parker, Q.A., Udalski, A. 2009a, A\&A, 496, 813	\\	\\
Miszalski, B., Acker, A., Parker, Q.A., \& Moffat, A.F.J. 2009b, A\&A, 505, 249 	\\	\\
Moe M., De Marco, O. 2006, ApJ, 650, 916	\\	\\
Morgan, D.H., Parker, Q.A., Cohen, M. 2003, MNRAS, 346, 719	\\	\\
Munari, U., Zwitter, T. 2002, A\&A, 383, 188	\\	\\
Neckel, T., Staude, H.J. 1984, A\&A, 131, 200	\\	\\
Nordstr\"om, B. 1975,  A\&AS, 21, 193	\\	\\
Nussbaumer, H. 1996, Ap\&SS, 238, 125	\\	\\
Ortiz, R., Lorenz-Martins, S., Maciel, W.J., Rangel, E.M. 2005,  A\&A, 431, 565	\\	\\
Pakull, M.W., 2009, in IAU Symp. 256, The Magellanic System: Stars, Gas, and Galaxies, ed. J.T van Loon, J.M. Oliveira  (Cambridge: CUP), p. 437	\\	\\
Pakull, M.W., Angebault, L.P. 1986, Nature, 322, 511	\\	\\
Parker, Q.A., Morgan, D.H. 2003, MNRAS, 341, 961	\\	\\
Parker, Q.A. et al. 2003, in ASP Conf. Ser. 209, Planetary Nebulae and their Role in the Universe, ed.  M. Dopita, S. Kwok, R. Sutherland  (San Francisco: Astron. Soc. Pacific), p. 41 
Parker, Q.A. et al. 2005, MNRAS, 362, 689	\\	\\
Parker, Q.A. et al. 2006, MNRAS, 373, 79	\\	\\
Parsamyan, E.S., Petrosyan, V.M. 1979,  Soobsch. Byurakan Obs., 51, 1	\\	\\
Payne, J.L., White, G.L., Filipovi\'c, M.D. 2008, MNRAS, 383, 1175	\\	\\
Pe\~na, M., Peimbert, M., Torres-Peimbert, S., Ruiz, M.T., Maza, J. 1995, ApJ, 441, 343	\\	\\
Perek L., Kohoutek L. 1967, Catalogue of Galactic Planetary Nebulae (Prague: Academia Publishing House)	\\	\\
Pereira, C.B., Miranda, L.F. 2005, A\&A, 433, 579	\\	\\
Pereira, C.B., Landaberry, S.J.C., de Ara\'ujo, F.X. 2003, A\&A, 402, 693	\\	\\
Pereira, C.B., Smith, V.V., Cunha, K. 2005, A\&A, 429, 993	\\	\\
Pereira, C.B., Marcolino, W.L.F., Machado, M., de Ara\'ujo, F.X. 2008, A\&A, 477, 877	\\	\\
Perinotto, M., Corradi, R.L.M. 1998, A\&A, 332, 721	\\	\\
Persson, S.E. 1988, PASP, 100, 710	\\	\\
Phillips, J.P., Cuesta, L. 1999, AJ, 118, 2919	\\	\\
Phillips, J.P., Guzman, V. 1998, A\&AS, 130, 465	\\	\\
Phillips, J.P., Ramos-Larios, G. 2008, MNRAS, 386, 995	\\	\\
Pierce M.J., Frew D.J., Parker Q.A., K\"oppen J. 2004,   PASA, 21, 334	\\	\\
Podsiadlowski, P., Morris, T.S., Ivanova, N. 2006, in ASP Conf. Series, 355, Stars with the B[e] Phenomenon, ed. M. Kraus, A.S. Miroshnichenko (San Francisco: ASP), p. 259	\\	\\
Poelarends, A.J.T., Herwig, F., Langer, N., Heger, A. 2008, ApJ, 675, 614	\\	\\
Polcaro, V.F. 2006, in ASP Conf. Series, 355, Stars with the B[e] Phenomenon, ed. M. Kraus, A.S. Miroshnichenko (San Francisco: ASP), p. 309	\\	\\
Porter, J.M., Rivinius T. 2003, PASP, 115, 1153	\\	\\
Pottasch, S.R. 1992, A\&ARev, 4, 215	\\	\\
Pottasch, S.R., Olling, R., Bignell, C., Zijlstra, A.A. 1988, A\&A, 205, 248	\\	\\
Preite-Martinez, A. 1988, A\&AS, 76, 317	\\	\\
Prieto, J.L., Sellgren, K., Todd, A. Thompson, T.A., Kochanek, C.S. 2009, ApJ, 705, 1425	\\	\\
Proga, D., Mikolajewska., J., Kenyon, S.J. 1994, MNRAS, 268, 213	\\	\\
Raga, A.C., B\"ohm, K.-H., Cant\'o, J. 1996, RMxAA, 32, 161	\\	\\
Raga, A.C., Riera, A., Mellema, G., Esquivel, A., Vel\'azquez, P.F. 2008, A\&A, 489, 1141	\\	\\
Ramos-Larios, G., Guerrero, M.A., Su\'arez, O., Miranda, L.F.; G\'omez, J.F. 2009, A\&A, 501, 1207	\\	\\
Rappaport, S., Chiang, E., Kallman, T., Malina, R. 1994, ApJ, 431, 237	\\	\\
Ratag, M.A., Pottasch, S.R., Zijlstra, A.A., Menzies, J. 1990, A\&A, 233, 181	\\	\\
Rauch, T., K\"oppen, J., Werner, K. 1994, A\&A, 286, 543	\\	\\
Rauch, T., Dreizler, S., Wolff, B. 1998,  A\&A, 338, 651	\\	\\
Rauch, T., Reiff, E., Werner, K., Kruk, J.W. 2008, in ASP Conf.  Ser., 391, Hydrogen-Deficient Stars, ed. K. Werner, T. Rauch (San Francisco: ASP), p.135	\\	\\
Reid, W.A., Parker, Q.A. 2006a, MNRAS, 365, 401	\\	\\
Reid, W.A., Parker, Q.A. 2006b, MNRAS, 373, 521	\\	\\
Reipurth, B., Bally, J. 2001, ARA\&A, 39, 403	\\	\\
Remillard, R.A., Rappaport, S., Macri, L.M. 1995, ApJ, 439, 646	\\	\\
Riera, A., Phillips, J.P., Mampaso, A. 1990, Ap\&SS, 171, 231	\\	\\
Riera, A., Garc\'ia-Lario, P., Manchado, A., Pottasch, S.R., Raga, A.C. 1995, A\&A, 302, 137	\\	\\
Riesgo-Tirado, H., L\'opez, J.A., 2002, RMxAA(SC), 12, 174	\\	\\
Riesgo, H., L\'opez, J.A. 2006, RMxAA, 42, 47	\\	\\
Rodr\'iguez, M., Corradi, R.L.M., Mampaso, A. 2001, A\&A, 377, 1042	\\	\\
Rosado, M., Kwitter, K.B. 1982, RMxAA, 5, 217	\\	\\
Sabbadin, F., Bianchini, A. 1979, PASP, 91, 278	\\	\\
Sabbadin, F., Hamzaoglu, E. 1981,  A\&A, 94, 25	\\	\\
Sabbadin, F., Minello, S., Bianchini, A. 1977,  A\&A, 60, 147	\\	\\
Sabin, L. et al. 2010, PASA, in press (these proceedings)	\\	\\
Sahai, R., Morris M., S\'anchez Contreras, C., Claussen, M. 2007, AJ, 134, 2200	\\	\\
Santander-Garc\'ia, M., Corradi, R.L.M., Mampaso, A., Morrisset, C., Munari, U., Schirmer, M., Balick, B., Livio, M. 2008, A\&A, 485, 117	\\	\\
Santander-Garc\'ia, M., Corradi, R.L.M., Mampaso, A., 2009, in Asymmetrical Planetary Nebulae IV, IAC Electronic Pub., ed. R.L.M. Corradi, A. Manchado, N. Soker (La Palma: Instituto de Astrof\'isica de Canarias), p. 555	\\	\\
Schmeja, D., Kimeswenger, S. 2001,  A\&A, 377, L18	\\	\\
Schmid, H.M. 1989,  A\&A 221, L31	\\	\\
Schmid, H.M., Nussbaumer, H. 1993, A\&A, 268, 159	\\	\\
Schwarz, H.E. 1991, A\&A, 243, 469	\\	\\
Schwarz, H.E., Corradi, R.L.M. 1992, A\&A, 265, L37	\\	\\
Schwarz, H.E., Corradi, R.L.M., Melnick, J. 1992, A\&AS, 96, 23	\\	\\
Schwarz, H.E., Aspin, C., Corradi, R.L.M., Reipurth, B. 1997,  A\&A 319, 267 	\\	\\
Shara, M., Moffat, A.F.J., Webbink, R.F. 1985, ApJ, 294, 271	\\	\\
Shara, M.M. 2007, Nature, 446, 159	\\	\\
Shaw, R.A., Rest, A., Damke, G., Smith, R.C., Reid, W.A., Parker, Q.A. 2007,  ApJ, 669, L25	\\	\\
Shull, C.F., McKee, J.M. 1979,  ApJ, 227, 131	\\	\\
Siviero, A. et al. 2007,  Baltic Astron., 16, 52	\\	\\
Smith, N. 2003, MNRAS, 342, 383  	\\	\\
Smith, N. 2006, ApJ, 644, 1151	\\	\\
Smith, N. 2007, AJ, 133, 1034 	\\	\\
Smith, N. 2009, in Asymmetrical Planetary Nebulae IV, IAC Electronic Pub., ed. R.L.M. Corradi, A. Manchado, N. Soker (La Palma: Instituto de Astrof\'isica de Canarias), p. 561 	\\	\\
Smith, N., Gehrz, R.D. 2005, AJ, 129, 969	\\	\\
Smith, N., Morse, J.A. 2004, ApJ, 605, 854	\\	\\
Smith, N., Bally, J., Walawender, J. 2007, AJ, 134, 846	\\	\\
Smith, N. et al. 2009, ApJ, 697, L49 	\\	\\
Soker N. 1997, ApJS, 112, 487	\\	\\
Soker N. 1998, ApJ, 496, 833	\\	\\
Soker N. 2002, A\&A, 386, 885 	\\	\\
Soker N., Subag, E. 2005, AJ, 130, 2717 	\\	\\
Soker N., Livio, M. 1989, ApJ, 339, 268	\\	\\
Soker, N., Tylenda, R. 2003, ApJ, 582, L105	\\	\\
Stahl, O. 1987, A\&A, 182, 229	\\	\\
Stupar, M., Parker, Q.A.,  Filipovi\'c, M.D. 2008, MNRAS, 390, 1037	\\	\\
Su\'arez, O., Garc\'ia-Lario, P., Manchado, A., Manteiga, M., Ulla, A., Pottasch, S.R. 2006, A\&A, 458, 173	\\	\\
Szczerba, R., Si\'odmiak, N., Stasi\'nska, G., Borkowski, J. 2007, A\&A, 469, 799	\\	\\
Tajitsu, A., Tamura, S., Yadoumaru, Y.,  Weinberger, R. \& K\"oppen, J. 1999,  PASP, 111, 1157	\\	\\
Thackeray, A.D. 1950,  MNRAS, 110, 524	\\	\\
Torres-Peimbert, S., Dufour, R.J., Peimbert, M., Pe\~na, M. 1997, in Planetary Nebulae, IAU Symp.  180, ed. H.J. Habing, H.J.G.L.M. Lamers (Dordrecht: Kluwer), 281	\\	\\
Tuthill, P.G., Lloyd, J.P. 2007, Science, 316, 247	\\	\\
Van de Steene, G.C., Pottasch, S.R. 1993, A\&A, 274, 895	\\	\\
Van der Veen, W.E.C.J., Habing, H.J. 1988, A\&A, 194, 125	\\	\\
Van der Veen, W.E.C.J., Habing, H.J., Geballe, T.R. 1989, A\&A, 226, 108	\\	\\
Van Winckel, H. 2003, ARA\&A, 41, 391	\\	\\
Van Winckel, H., Schwarz, H.E., Duerbeck, H.W., Fuhrmann, B. 1994, A\&A, 285, 241	\\	\\
Vaytet, N.M.H. et al., 2009, MNRAS, 398, 385	\\	\\
Veilleux, S., Osterbrock, D.E. 1987, ApJS, 63, 295	\\	\\
Viironen, K., Delgado-Inglada, G., Mampaso, A., Magrini, L., Corradi, R.L.M. 2007, MNRAS, 381, 1719	\\	\\
Viironen, K. et al. 2009a, A\&A, 502, 113 	\\	\\
Viironen, K. et al. 2009b, A\&A, 504, 291  	\\	\\
Wareing, C.J. 2010, PASA, in press (this issue), eprint: arXiv:0910.2200	\\	\\
Webster, B.L. 1978, MNRAS, 185, 45P	\\	\\
Weinberger, R., Tajitsu, A., Tamura, S., Yadoumaru, Y. 1998, PASP, 110, 722	\\	\\
Werner, K., Herwig, F. 2006, PASP, 118, 183	\\	\\
Wesson, R., Barlow, M.J., Liu, X.-W., Storey, P.J., Ercolano, B., De Marco, O. 2008a, MNRAS, 383, 1639	\\	\\
Wesson, R. et al. 2008b, ApJ, 688, L21	\\	\\
Whitelock, P.A., Menzies, J.W. 1986, MNRAS, 223, 497	\\	\\
Whiting, A.B., Hau, G.K.T., Irwin, M. 2002, ApJS 141, 123	\\	\\
Whiting, A.B., Hau, G.K.T., Irwin, M., Verdugo, M. 2007, AJ, 133, 713	\\	\\
Wray, J.D. 1966, Unpublished PhD thesis, Northwestern University	\\	\\
Zanin, C., Kerber, F. 2000, A\&A, 356, 274	\\	\\
Zickgraf, F.-J. 2003, A\&A, 408, 257	\\	\\
Zijlstra, A.A. 2002, Ap\&SS, 279, 171	\\	\\
Zijlstra, A.A., Pottasch, S.R. 1991, A\&A, 243, 478	\\	\\

\end{document}